\documentstyle[aps,epsf,amsfonts,psfrag,amsmath]{revtex}

\newcommand{\vp}{\varphi}

\newcommand{\k}{l}

\begin{document}
\date{\today}

\title{Moduli stabilization in higher dimensional brane  models}

\author{Antonino Flachi$^1$\footnote{flachi@ifae.es}, Jaume
  Garriga$^{1,2}$\footnote{garriga@ifae.es}, Oriol
  Pujol{\`a}s$^1$\footnote{pujolas@ifae.es} and Takahiro
  Tanaka$^{3,4}$\footnote{tanaka@yukawa.kyoto-u.ac.jp}}

\author{$^{}$}

\address{$^1$ IFAE, Campus UAB,
  08193 Bellaterra (Barcelona), Spain}

\address{$^2$ Departament de F{\'\i}sica Fonamental and C.E.R. en Astrof{\'\i}sica, F{\'\i}sica
  de Part{\'\i}cules i Cosmologia, \\Universitat de Barcelona, Mart{\'\i} i Franqu{\`e}s 1,
08028 Barcelona, Spain}

\address{$^3$ Institute of Cosmology, Department of Physics and Astronomy, Tufts
  University, Medford MA 02155, USA}

\address{$^4$ Yukawa Institute for Theoretical Physics, Kyoto University, Kyoto
  606-8502, Japan}

\maketitle

\begin{abstract}

    We consider a class of warped higher dimensional brane models with topology
    ${\cal M} \times \Sigma \times S^1/Z_2$, where $\Sigma$ is a
    $D_2$ dimensional manifold. Two branes of codimension one are
    embedded in such a bulk space-time and sit at the orbifold fixed points.  We
    concentrate on the case where an exponential warp factor
    (depending on the distance along the orbifold) accompanies the
    Minkowski ${\cal M}$ and the internal space $\Sigma$ line
    elements. We evaluate the moduli effective potential induced by bulk scalar
    fields in these models, and we show that generically this can stabilize the
    size of the extra dimensions. As an application, we consider a scenario where
    supersymmetry is broken not far below the
    cutoff scale, and the hierarchy between the electroweak and the
    effective Planck scales is generated by a combination of redshift and large
    volume effects.  The latter is efficient due to the shrinking of $\Sigma$ at
    the negative tension brane, where matter is placed. In this case, we find that
    the effective potential can stabilize the size of the extra dimensions (and the
    hierarchy) without fine tuning, provided that the internal space $\Sigma$ is flat.
\end{abstract}

\hfill \vbox{ \hbox{YITP-01-85} \hbox{UAB-FT-539}}

\newcommand{\nn}{\nonumber\\}
\newcommand{\hsi}{\hat{\sigma}}
\newcommand{\hrho}{\hat{\rho}}
\newcommand{\hl}{\hat{\lambda}_\k}
\newcommand{\si}{\sigma}
\newcommand{\dlt}{\bigtriangleup}
\newcommand{\beq}{\begin{equation}}
\newcommand{\eeq}{\end{equation}}
\newcommand{\bed}{\begin{displaymath}}
\newcommand{\eed}{\end{displaymath}}
\def\bea{\begin{eqnarray}}
\def\eea{\end{eqnarray}}

\section{Introduction}
\label{introduction} {}

It is a widely accepted idea that our space-time might have more
than four dimensions. This very old proposal \cite{Ka,Kl}, revived
in \cite{aadd}, is currently being considered in several contexts
from particle physics to cosmology.

From a more fundamental perspective, string theory provides a
strong motivation for considering higher dimensional scenarios.
In a suggestive model \cite{hw,stelle}, Ho\v rava and Witten
considered an $11-$dimensional supergravity theory compactified on
${\cal M} \times {\cal CY} \times S^1/Z_2$, where ${\cal M}$
represents four dimensional Minkowski space and ${\cal CY}$ is a
Calabi-Yau space, with the size of the Calabi-Yau much smaller
than the orbifold one. In such a set-up, the fixed points of the
$Z_2$ symmetry can be viewed as $10-$dimensional boundaries which
may accommodate a realistic gauge theory.

A closely related model has been considered by Randall and Sundrum
\cite{RS1}, where two flat $3-$branes of opposite tension are
embedded in $5-$dimensional Anti-de Sitter space. The extra
dimension is compactified on a $Z_2$ orbifold, with the branes
sitting at the fixed points. This results in a space-time with a
non-factorisable geometry which, through a redshift effect,
induces a four-dimensional effective scale on one of the branes
which is much smaller than the $5-$ dimensional cutoff scale
(which is assumed to be  Planckian). This geometric effect could
explain the hierarchy between the Planck and the electroweak
scales. The "radion" mode (which corresponds to the size of the
extra dimension) plays a crucial role in this scenario, since its
vacuum expectation value (VEV) determines the hierarchy. In the
original RS model, the radion was massless at the classical level.
This would cause an unacceptable modification of the resulting
gravity on the brane \cite{gt}, and therefore the radion must be
stabilized by some mechanism \cite{gw}. Moreover, for a
satisfactory solution to the hierarchy problem, such mechanism
should give the radion a suitably large VEV without the need of
introducing any fine tuning of parameters.

The problem of radion stabilization is not particular to the RS
model, and arises generically in higher dimensional Kaluza-Klein
theories, including the usual factorisable geometries based on
direct products of Minkowski space with homogeneous manifolds. In
that context, Candelas and Weinberg realized that the Casimir
energy of matter fields or gravity could induce an effective
potential for the radion at the quantum level, offering a natural
mechanism for stabilizing the size of the extra dimensions
\cite{cw}. Following this idea, a number of authors have
investigated quantum stabilization in the context of brane models.
In \cite{flachi1,flachi2,flachi3,pujolas1,pujolas2,toms,roth1} the
lowest order quantum corrections arising from bulk fields on the
RS background have been calculated and their relevance to the
radius stabilization problem has been investigated (see also
\cite{sasaki,sasaki2,wade,nojiri1,ogushi,mukohyama} for
discussions of the possible relevance of quantum effects in
cosmological brane-world scenarios and \cite{brevik} for finite
temperature effects).  Initially, such investigations were
somewhat discouraging, since it was found \cite{pujolas1} that in
the RS model it is not possible to stabilize the extra dimension
and simultaneously solve the hierarchy problem by means of the
Casimir forces due to bulk gravitons or generic bulk scalars.
However, it has recently been shown \cite{gapo} that bulk gauge
fields (or any of their supersymmetric relatives) can do the job.
These fields induce logarithmic contributions to the radion
effective potential which are sufficient to stabilize it,
generating a large hierarchy of scales without fine-tuning. The
logarithmic behaviour can be understood, in a 4D holographic
description, as the running of gauge couplings with the infrared
cut-off scale (which corresponds to the electroweak scale).

The RS model has opened up a very interesting framework for model
building in particle physics, with possible cosmological
implications. This scenario, however, is just the simplest
possibility within a more general class of higher dimensional
warped geometries which deserve fuller exploration, specially with
regard to their possible embedding in string theory. In this
connection, one expects that more general internal spaces and
higher dimensionalities have to be considered,
and in this paper we shall take a step in this direction. Our aim
is twofold. On one hand, the consideration of more general
spacetimes may provide interesting extensions of the RS mechanism
for the geometric origin of the hierarchy. On the other, quantum
effects in such scenarios can be qualitatively different,
providing new ways of stabilizing the radion which do not
necessarily rely on the peculiar behaviour of bulk gauge fields.

Generically, we expect that the behaviour of the effective
potential for the "moduli" should be qualitatively different once
we go beyond the RS scenario. This is indicated (even in five
dimensions) by the study of generalized warped compactifications
\cite{pujolas2}. Aside from the non-local Casimir interaction
between the branes which we mentioned above, local terms which are
induced by quantum effects may stabilize the moduli when we
consider warped brane worlds where the bulk is different from AdS.
In the RS model both the branes and the bulk space-time are
maximally symmetric and thus any possible counter-term amounts to
a renormalization of the brane tensions. However, this is not true
in general. An explicit example was worked out in detail in
\cite{pujolas2}, where a class of $5-$dimensional brane models
with power law warp factors were investigated. In this case, the
global symmetry which is responsible for the masslessness of the
moduli at the classical level is shown to be anomalous. Thus, the
effective potential develops terms which do not scale
appropriately under the global symmetry and which therefore act as
stabilizers for the moduli. Some of the 5 dimensional models
considered in Ref. \cite{pujolas2} can be obtained by dimensional
reduction of $5+D_2$-dimensional models, and in this paper we
shall focus on a class of higher dimensional models which includes
those.

Specifically, we shall consider spaces with line element given by
 \begin{equation}
    \label{metricansatz}
    ds_{(D+1)}^2= e^{2\sigma(y)}\eta_{\mu\nu}dx^\mu dx^\nu
    + R^2 e^{2\rho(y)} \gamma_{ij}dX^idX^j + dy^2.
\end{equation}
The coordinates $x^\mu$ parametrize four dimensional Minkowski
space ${\cal M}$ and the coordinates $X^i$ cover a
$D_2-$dimensional compact internal manifold $\Sigma$. We locate
two $D\equiv(4+D_2)-$dimensional branes at the fixed points of the
orbifolded dimension labelled by $y$.
Such a metric is found as a solution of a $(D+1)-$dimensional system of gravity
plus certain 'matter' fields. Depending on the field content, different warpings
can arise. For instance, Gregory showed in \cite{gregory} that a six dimensional
global string solution exists for negative cosmological constant, with
$\sigma(y)=-k|y|$ and $\rho(y)=constant$. Gherghetta and Shaposhnikov \cite{gherghetta}
constructed the metric solution of a six dimensional local string-like defect
with $\sigma (y) = \rho (y)= -k|y|$. Generalizations of this models with more
extra dimensions with $\Sigma = S^{D_2}$ were considered in in \cite{ghrsh,oda}
using bulk scalar fields with a hedgehog configuration.

The authors of \cite{daemi} include Yang-Mills (YM) fields with appropriately chosen
gauge group, instead. They find a series of solutions classified in terms of the
Ricci flatness of the manifold $\Sigma$: when the internal space is Ricci flat
(for example a $D_2-$dimensional torus or a Calabi-Yau space), one obtains warp
factors which can generically be expressed as sums of exponentials if there is
no YM flux; in particular, when the bulk cosmological constant is negative, a
specially simple solution with both warp factors equal to the RS one exists
(including the case of higher dimensional $AdS$ space).  Turning on some YM flux
can relax the condition of the Ricci flatness of the internal manifold. 
In this case, they find a
solution where along the Minkowski direction the warp is {\em {\`a} la} RS, whereas
along the curved manifold it is constant. This gives the interesting combination
of a higher dimensional theory which is a hybrid between an ordinary
Kaluza-Klein theory and the RS model. In passing, we note that the phenomenology
of such a scenario has been recently considered in \cite{rizzo,shap}.

In the present paper we start considering quantum effects from
bulk fields quantized on such space-times beginning with the case
when the two warp factors are of the
RS type. The computation is similar in some respect to the RS one
with some additional technical complications. Moreover we allow a
more general setting than the one presented in \cite{daemi} in
that we consider solutions with internal manifolds $\Sigma$ which
are not Ricci flat.
In  section \ref{setup} we describe   how to construct these types
of models
(for the case of Ricci flat
internal manifolds we refer the reader to reference \cite{daemi}) and
subsequently discuss the moduli approximation in this type of space-times.

The Kaluza-Klein reduction of the bulk field is considered in section
\ref{kkreduction}, where it is shown that, when branes have a non trivial
topology, the physical mass depends also on the eigenvalues of the Klein-Gordon
operator on the internal manifold.  The one-loop effective action is evaluated
in section \ref{qea} and the result is simply expressed in terms of
heat-kernel coefficients. This is done using the Mittag-Leffler expansion of the
generalized $\zeta$ function. The result is regulated by using dimensional
regularization and appropriately renormalized in section \ref{ren}.

In section \ref{scales}, we propose a scenario where
supersymmetry is broken just below the cutoff scale and the
hierarchy is generated by a combination of redshift and large
volume effects.
The possibility of
stabilizing the moduli (and the hierarchy) with the quantum effective potential is discussed in
section \ref{stab} for the different cases of interest and it is shown that
such a stabilization can in fact take place without fine tuning. The conclusions
are left to the last section.

The paper is equipped with an appendix in which we obtain the
Mittag-Leffler expansion for a positive semidefinite operator
(This last result does not claim any originality and can possibly
be found elsewhere). Appendix \ref{app:5dreduct} contains the
details of the dimensional reduction of the gravitational part of
the action.

\section{Setup}
\label{setup}

We are interested in the quantum effective action arising
from a quantized bulk scalar field on the background
space-time
specified by the metric (\ref{metricansatz}). We consider
two branes of codimension one with the topology of
${\cal M}^4 \times \Sigma$, where the manifold $\Sigma$ is
taken to be Einstein and compact.
The branes sit at the orbifold fixed points and the $Z_2$
symmetry is imposed on the solutions.

As we have already mentioned in the introduction, such solutions
have been obtained in \cite{daemi} for Ricci flat internal
manifolds. However, as also shown in \cite{ghrsh,oda}, more
general solutions with $\Sigma = S^{D_2}$ can be found by
introducing additional matter content coming from a scalar field
with a hedgehog configuration. In the next Section, we show how to
obtain such a solution.

\subsection{Model}
\label{model}

The specific space-time that we consider in this paper corresponds
to the case when the two warp factors are equal and exponential,
$\sigma(y)=\rho(y)=-k|y|$.  The model consists of an $G$ invariant
non-linear sigma model parametrized by a set of bulk scalar fields
$\phi^a$ together with a standard bulk gravity sector, and two
boundary-branes. This is described by the action
\begin{eqnarray}
    \label{modelaction}
    S&=&\int d^{D+1}x \sqrt{g_{(D+1)}} \Biggl\{ M^{D-1}
    {\cal R}_{(D+1)} - \Lambda
    - \partial_M\phi^{a\dagger} \partial^M\phi^a
    - \lambda (\phi^{a\dagger}\phi^a -v^2)       \Biggr\}  \cr
    &-& \int d^{D}x  \sqrt{g_{(D)+}} \; \tau_+ - \int d^{D}x
    \sqrt{g_{(D)-}} \; \tau_-.
\end{eqnarray}
Our notation the following. The higher dimensional bulk
indices are $M, N,\dots$ and run over $\mu,i,y$; the $(4+D_2)\equiv D$
dimensional brane indices are $A,B,\dots$ and run over
$\mu,i$; $g^{(D+1)}_{MN}$ is the bulk metric and
$g^{(D)\pm}_{AB}$ are the induced metrics on the branes.
Finally, $\tau_\pm$ are the brane tensions, and $M$ is the
higher dimensional fundamental Planck mass.

Let us look more closely at the structure of the scalar
fields.  The equation of motion for the scalars can be
written as usual:
\begin{equation}
    \Box \phi^a = - \lambda \phi^a.
    \label{eomphi1}
\end{equation}
The role of non-dynamical auxiliary field $\lambda$ is to
impose the constraint
$$
\phi^{a\dagger}\phi^a-v^2=0.
$$
Differentiating this constraint twice, we can rewrite
Eq. (\ref{eomphi1}) as follows:
\begin{equation}
    \Box \phi^a = - \left({\partial_M\phi^b
\partial^M\phi^{b\dagger}\over v^2}\right) \phi^a.
    \label{eomphi2}
\end{equation}
The previous equation allows hedgehog solutions for $\phi^a$ for
suitable choices of the group $G$. Moreover, they have a constant
profile along the orbifold  and satisfy
\begin{equation}
    \label{hedgehog}
    \Delta_\gamma \phi^a=-L^2 \phi^a, \qquad\text{and}\qquad
    \partial_M \phi^{a\dagger} \partial^M \phi^a =
    e^{-2\rho} {L^2 v^2 \over R^2}
\end{equation}
where  $L$ is a 'winding number', and $\Delta_\gamma$ is the
laplacian obtained from $\gamma_{ij}$.

The Einstein equations for such hedgehog configurations have been
studied in \cite{gherghetta,ghrsh,oda}, where solutions of the
type (\ref{metricansatz}) with $\sigma(y)=\rho(y)= - k y$ have
been found, with
\begin{equation}
    \label{kLambda}
    k=\sqrt{-4 M^{1-D}\Lambda / (D-1)D},
\end{equation}
where $\Lambda<0$. In order to obtain the space-time described
previously we take two copies of a slice of this $D+1$ dimensional
space comprised between $y_+$ and $y_-$, corresponding to the
brane locations. The two copies are glued together there. Along
with the the identification $y-y_\pm\to 2y_\pm-y $, this gives the
topology of an $S^1/Z_2$ orbifold in the $y$ direction.

In order for this to be a solution of our model
(\ref{modelaction}), the brane
tensions have to satisfy
\begin{equation}
    \label{tensions}
    \tau_\pm=\pm4\sqrt{-(D-1)M^{D-1}\Lambda /
      D}=\pm4(D-1)M^{D-1} k,
\end{equation}
as a result of the junction conditions at the branes. Besides
(\ref{kLambda}), the Einstein equations in the bulk relate the
hedgehog parameters to the curvature of the internal manifold
$\Sigma$  as
\begin{equation}
    \label{sigmahedgehog}
    v^2 = {2D_2 C \over L^2} M^{D-1}.
\end{equation}
Here, since $\Sigma$ is homogeneous, the dimensionless constant $C$ is defined
through ${\mathcal R}^{(\gamma)}_{ij}=C\gamma_{ij}$, and ${\mathcal
  R}^{(\gamma)}_{ij}$ is the Ricci tensor computed out of $\gamma_{ij}$.

Associated to the sigma model scalars a number of Nambu Goldstone modes will be
present. However, we shall assume that these couple to matter only through
gravity, so that their effects are negligible.

\subsection{Moduli}
\label{moduli}

One interesting feature of this ansatz is that the
parameter $R$, describing
the volume of $\Sigma$, does not appear in the equations
of motion even in the
case of a curved internal space. Moreover, the positions of
both branes are free at
the classical level.  They correspond to flat directions in
the action and thus
are the relevant degrees of freedom at low energies. In the
moduli
approximation, which we shall follow here, they are promoted
to four dimensional
scalar fields.

One crucial difference of these solutions with respect to the RS
model is that they are not homogeneous along the orbifold, even in
the case when $\Sigma$ is a torus \footnote{In this case, the
solution corresponds to a toroidal compactification of a higher
dimensional $AdS$ space.}. This is due to the compactness of
$\Sigma$. In contrast with the RS model, the positions of both
branes are physically meaningful.

However, it is clear that a scaling of $R$ is equivalent to
a shift in the
positions of the branes $y_\pm$. Therefore, they are not
independent. Rather,
only two moduli are needed. Since we will use several
combinations of the moduli
along the paper, we summarize them briefly now:
$a_\pm\equiv e^{-k y_{_\pm}}$, the physical radii of
$\Sigma$ at the branes
$R_\pm=a_\pm R$, the corresponding dimensionless values
$r_\pm=a_\pm k R$, and
$a\equiv e^{-k(y_{_-}-y_{_+})}=a_-/a_+$.

%
%

In addition to the moduli, the massless sector also
contains the graviton zero
mode. To take it into account, we perturb the background
solution
(\ref{metricansatz}) as follows
\begin{equation}
    \label{pertbulkmetr}
    ds^2=dy^2+e^{2\sigma(y)}\left[\tilde g_{\mu\nu}(x)dx^\mu
dx^\nu + R^2 \gamma_{ij} dX^i
    dX^j\right].
\end{equation}
Substituting this metric back into the action (\ref{modelaction})
we obtain the kinetic term for $\tilde g$ coming from the bulk
part (see \cite{pujolas2}). The kinetic terms for the moduli
$y_\pm$ come from the boundary terms. A computation analogous to
that in \cite{pujolas2} gives
\begin{equation}
    S_{(4)}= - m_P^2 \int d^4 x \sqrt{-\tilde g}
    \left\{ \left[\vp_+^2-\vp_-^2\right]\ \tilde{\cal R}
        - 4{D-1\over D-2} \left[(\tilde\partial  \vp_+)^2
            - (\tilde\partial \vp_-)^2 \right]
    \right\},
    \label{modulikin}
\end{equation}
where $\vp_\pm^2=a_\pm^{D-2}=e^{-(D-2)k y_{{}_\pm}}$,
and the effective four dimensional Planck mass is
given by
\begin{equation}
    \label{planckmass}
    m_P^2 = {2\over D-2} v_{{}_\Sigma}  R^{D_2} M^{D-1}/k ~,
    \qquad {\rm with}\qquad
    v_{{}_\Sigma}=\int_\Sigma \sqrt{\gamma} d^{D_2} X.
\end{equation}
We note that the moduli $\vp_\pm$ 
are Brans-Dicke (BD) fields and in the frame defined by
$\tilde g_{\mu\nu}$,
the kinetic term for $\vp_+$ has the wrong sign. Introducing
the new variables
$\vp$ and $\psi$ \cite{pujolas2,ekpy},
$$
\vp_+=\vp\,\cosh \psi \quad {\rm and}\quad \vp_-=\vp\,\sinh
\psi,
$$
the Einstein frame is given
by $\hat g_{\mu\nu}=\vp^2 \tilde g_{\mu\nu}$. In this frame
the action
(\ref{modulikin}) takes the form
\begin{equation}
    S_{(4)}= - m_P^2 \int d^4 x \sqrt{-\hat g}
    \left\{ \hat{\cal R}
        + 2{D_2\over D_2+2} (\hat\partial  \ln\vp)^2
        + 4{D_2+3\over D_2+2} (\hat\partial \psi)^2
    \right\},
    \label{einsteinframe}
\end{equation}
and now the kinetic terms are both positive definite.
Moreover, we note that the
modulus $\vp$ decouples in the limit $D_2\to0$, as
expected, since this case
corresponds to the usual RS model, where only one modulus is
present.

We are assuming that the ($D$ dimensional) matter
fields
$\chi^{(D)}_\pm$ are localized on each brane and
so they couple universally to the
corresponding induced metrics
$g^{(D)\,\pm}_{AB}$ (recall $A,B,\dots=\mu,i$)
\begin{equation}
    \label{matteraction}
    S^{\rm{matt}}=\sum_{\pm}\int d^{D}x\,
    \sqrt{-g^{(D)\pm}}\,
    {\cal L}^{\pm}\left(\chi^\pm_{(D)},
    g^{(D)\,\pm}_{AB}\right)
    =\int  d^{4}x
    \sum_{\pm} \, \sqrt{-g_{\pm}} \,
    a_\pm^{D_2} {\cal L}^{\pm}\left(\chi^\pm,
g^{\pm}_{\mu\nu}\right).
\end{equation}
Here, we have kept the $\Sigma -$zero modes only,
and integrated out the $X$
dependence,  the $\Sigma$ volume factor has been
absorbed by the four dimensional
matter fields $\chi$ and couplings, and the four dimensional
induced metrics are the $(\mu,\nu)$ components of the $D$
dimensional ones $g^\pm_{\mu\nu}=g^{(D)\,\pm}_{\mu\nu}$.


%
%


A repeated use of  the chain rule leads to the interaction of the moduli with
matter given by
\begin{equation}
    \label{modmatteinst}
    S^{\rm mod-matt} =
    \int d^4x\, \sqrt{-\hat g}  \Biggl\{
    -{D_2\over D_2+2}
    \sum_\pm \left[\hat{ T}_\pm -2 \hat{\cal L}_\pm
    \right]    \delta \ln \vp
    +{2\over D_2+2}
    \sum_\pm a^{\pm (D_2+2)/2} \left[ \hat{T}_\pm +D_2 \hat{\cal L}_\pm
    \right] \delta    \psi
    \Biggr\},
\end{equation}
where $\hat {T}_\pm$ and $\hat {\cal L}_\pm $ are defined according to
$\sqrt{-g_\pm} \,{T}_\pm = \sqrt{-\hat g}\, \hat { T}_\pm$, and $\sqrt{-g_\pm}
a_\pm^{D_2}\,{\cal L}_\pm = \sqrt{-\hat g}\, \hat {\cal L}_\pm$.  The coupling
of the moduli to the Lagrangian is entirely due to the dimensions along $\Sigma$
being warped, and is a generic prediction of models with a nontrivial warp
factor for the extra dimensions. In fact, the 'radion' modulus $\Psi$ is coupled
to matter through $(\hat T+D_2 \hat {\cal L})_\pm$, which coincides with the trace of the
$D$ dimensional energy momentum tensor.
Moreover, this shows that the the modulus $\vp$ decouples from matter in the RS
limit $D_2\to0$, as it should.

Defining the canonical fields
$$
\Phi = 2\sqrt{D_2\over D_2+2}\,m_P \;\delta \ln \vp, \qquad {\rm
and}\qquad
\Psi = 2\sqrt{2{D_2+3\over D_2+2}} \,m_P \; \delta\psi,
$$
we obtain the equations of motion for the moduli
\begin{eqnarray}
    \label{eommoduli}
    \hat \Box \Phi &=& {1\over2} \sqrt{D_2\over D_2+2}\,{1\over m_P}\,
    \left[ \hat T_+ -2 \hat {\cal L}_+ + \hat T_- -2 \hat {\cal L}_-\right] \cr
    \hat \Box \Psi &=& -{1\over\sqrt{2(D_2+3)(D_2+2)}\,
      m_P} \;\left[ a^{(D_2+2)/2} \left( \hat T_++D_2\hat{\cal L}_+ \right)
        + a^{-(D_2+2)/2} \left( \hat T_-+D_2\hat{\cal L}_- \right) \right].
\end{eqnarray}
As we explain in Sec. \ref{scales}, we are interested in the case of $a\ll1$ in
order to have a substantial redshift effect arising from the warp factors.
Unless otherwise stated, we shall set $\langle a_+\rangle =1$, so that, with a
good accuracy,
$ a_-\simeq  a$,
$\vp\simeq\vp_+\simeq 1$ and
$\psi\simeq\vp_-\ll 1$.

Thus, from (\ref{eommoduli}) we can read off  the couplings to the two types of
matter:\footnote{In the rest of this article, we will consider only matter
  located on the negative tension brane. Here we just consider other possible
  forms of matter at $y=y_+$ for the sake of generality.}
$\Phi$ 
couples to the matter at either brane $\chi_\pm$, with a strength $\sim 1/m_P$.
As for $\Psi$, 
the coupling to $\chi_-$ is quite large, of order $a^{-(D_2+2)/2}/m_P$, and to
$\chi_+$ is even smaller than Planckian,  $\sim a^{(D_2+2)/2}/m_P$.

\section{Kaluza-Klein Reduction}
\label{kkreduction}

Before the evaluation of the one-loop effective action, which we do in Section
\ref{qea}, we turn now to the reduction in KK modes of a bulk scalar field
living in the space-time described in the previous section.


The idea is very simple: by performing a Kaluza-Klein
reduction of the higher
dimensional scalar field theory from $D+1$ (with $D=D_1+D_2$) down to
$D_1$ dimensions, we
obtain an equivalent lower dimensional theory consisting of
an infinite
number of massive Kaluza-Klein modes.  Specifically, the
Kaluza-Klein reduction
is performed by expanding the higher dimensional scalar
field in terms of a
complete and orthogonal set of modes and then integrating
out the dependence on
the extra dimensions.  The masses turn out to be quantized
according to some
eigenvalue problem and depend on the details of the
space-time, the nature of the
internal manifold and on the bulk (higher dimensional)
scalar field. The one-loop
effective action can then be evaluated by re-summing the
contribution of each one
of the modes.

Typically in Kaluza-Klein theory the mass eigenvalues are
found explicitly and the subsequent
evaluation of the sum over the modes does not present
particular difficulties. However, in the case of
warped space-times the main difference is that the orbifold
nature of the extra dimension
complicates the mass eigenvalues, which are expressed in
terms of a transcendental equation
and thus cannot be found explicitly.

In the present section we will carry out the first step of
the computation, namely the Kaluza-Klein
reduction of the bulk scalar field. We will consider the
most general case of a massive non-minimally
coupled scalar field and assume that $\Sigma$, a compact manifold.

The bulk scalar field $\Upsilon(X,x,y)$ obeys the following
equation of motion:
\begin{equation}
    \left[-\Box_{(D+1)} +m^2+ \xi {\cal R}_{(D+1)}\right]
    \Upsilon =0~,
\label{eq3}
\end{equation}
where ${\cal R}_{(D+1)}$ is the higher dimensional curvature and
$\Box_{(D+1)}$ is the D'Alembertian, both computed
from the metric (\ref{metricansatz}).

Using the explicit expression for the metric tensor, we
can disentangle, in equation (\ref{eq3}), the
dependence on the internal manifold from the Minkowskian
one.
A straightforward calculation gives:
\bea
\Bigl[ -e^{-2 \sigma} \Box_{} -
e^{-2\rho}{1\over R^2} \Delta_{(\gamma)} -
e^{-{} \tau}
\partial_y e^{{} \tau} \partial_y &+&\nn
+ m^2 + \xi e^{-2\rho}
{1\over R^2}{\mathcal R}^{(\gamma)} - \xi F(y)
&\Bigr]& \Upsilon =0~,
\label{eq4}
\eea
where $\Delta_{(\gamma)}$ is the laplacian related  to
$\gamma_{ij}$, $\Box_{}$ is the $D_1$ dimensional flat
D'Alembertian, and
\begin{eqnarray}
    \label{eq5}
    F(y)&=& 2 \tau''(y) +
    {} \tau'(y)^2 +
    {} D_1 \sigma'(y)^2
        + D_2 \rho'(y)^2 ~, \nn
    \tau(y) &=& D_1 \sigma (y) +  D_2 \rho (y)~.
\end{eqnarray}
We now expand the field $\Upsilon(x,X,y)$ in terms of a
complete set of modes carrying a momentum along the orbifold
and $\Sigma$ directions labelled by indexes $n$ and $\k$
respectively,
\beq
\Upsilon(x,X,y) = \sum_{\k,n} \Psi_\k(X)
\Phi_{\k,n}(x)Z_{\k,n}(y).
\label{eq6}
\eeq
Here, the modes $\Psi_\k(X)$  are a complete set of
solutions of
the Klein-Gordon equation on the manifold $\Sigma$:
\beq
   P_\Sigma \Psi_\k(X) \equiv {1\over R^2}
   \Big[- \Delta_{(\gamma)} + \xi
   {\mathcal R}_{(\gamma)}\Big] \Psi_\k(X) =
   \lambda_\k^2
   \Psi_\k(X)~,
   \label{psigma}
\eeq
with eigenvalues $\lambda_\k^2$ and
degeneracy\footnote{Although we assume $P_\Sigma$ to be
either positive semidefinite or positive definite,
  the label $\k=0$ always refers to the zero eigenvalue,
i.e., $\lambda_0=0$, the
  existence of this eigenvalue being set by $g_0$ being $0$
or $1$.} $g_\k$.
If we now require $\Phi_{\k,n}(x)$ to satisfy the
Klein-Gordon equation on
the Minkowskian factor of the space-time $\mathcal M$ with
masses $m^2_{\k,n}$,
\beq
  \left[- \Box + m^2_{\k,n}\right]
  \Phi_{\k,n}(x)=0~,
\label{eq8}
\eeq
we are left with a radial equation for the modes
$Z_{\k,n}(y)$ of the form
\beq
    e^{2\sigma}\left[
    -e^{-\tau} \partial_y
    e^{{}\tau}\partial_y
    + m^2  - \xi F(y)
    +\lambda_\k^2 e^{-2\rho}
    \right] Z_{\k,n}
    =m^2_{\k,n} Z_{\k,n}~.
    \label{eq9}
\eeq
This equation is valid for any warp factors $\sigma$ and
$\rho$, and can be viewed as an eigenvalue problem for the
orbifold modes $Z_{\k,n}$ and the physical masses
$m_{\k,n}$. Both of them depend in general on the 'internal'
index $\k$.
In this paper we consider
the case of two equal warp factors, with
\bed
\rho(y)=\sigma(y)=-k|y|~.
\eed
Defining $D=D_1+D_2~$, we can
specialize Eq. (\ref{eq9}) to this case as
\beq
    \left[ -e^{(2-D)\si} \partial_y e^{D\si} \partial_y  +
        m^2 e^{2\si}  - \xi F(y) e^{2\si} \right] Z_{\k,n} =
    (m_{\k,n}^2 - \lambda_\k^2) Z_{\k,n} .
    \label{eq10}
\eeq
We note that the operator in the { l.h.s.} {\em does not}
depend on the internal index $\k$. Accordingly, in this case
neither  the modes $Z_{\k,n}$ nor the combination
$q^2_n\equiv m_{\k,n}^2-\lambda_\k^2$ depend on $\k$.
In other words, the dependence on $\k$ and $n$ of the masses
is factorized for this geometry,
\begin{equation}
    \label{kkmasses}
    m_{\k,n}^2=q_n^2+\lambda_\k^2.
\end{equation}
Therefore, from now on we shall drop this index in $Z$.
On the other hand, Eq. (\ref{eq10})  is
similar to the one which arises in the
RS model,
and the most general solution
can still be written in terms of Bessel
functions:
\beq
Z_{n}^\beta (y) = \epsilon_\beta (y)
\left[
A^\beta_{n} J_\nu \left({q_{n}\over k}e^{-\si}\right) +
B^\beta_{n} Y_\nu \left({q_{n}\over k}e^{-\si}\right)
\right]
\label{eq11}
\eeq
where for notational convenience we have defined
\begin{equation}
\label{eq12}
\epsilon_\beta (y) = e^{-{D\over 2}\sigma(y)}
\left\{
    \begin{minipage}[c]{5cm}
        $y/|y|\qquad$   $\beta =$ twisted \\
        $1~~\qquad$       $\beta =$ untwisted,
    \end{minipage}
\right.
\end{equation}
and
\beq
\nu^2 = {m^2 \over k^2} - D(D+1) \xi +{D^2 \over 4}~.
\label{eq13}
\eeq
The index $\beta$ has been introduced in order to
discriminate  the two possible cases of $\Upsilon$ being
untwisted    ($Z_{n}(-y) = Z_{n}(y)$) or twisted
($Z_{n}(-y) = -Z_{n}(y)$).
Imposing the appropriate boundary conditions, which can
be obtained by
integrating equation (\ref{eq10}) across the orbifold fixed
points, we find that the eigenvalues $q_{n}$ are
determined by the  transcendental equation:
\beq
F_\nu^\beta\left({q_{n}\over k a}\right)=0~,
\label{masses}
\eeq
where
\bea
\label{fs}
F_\nu^{\beta}(z) =  \left\{
    \begin{minipage}[l]{10cm}
        $ Y_\nu (az)J_\nu (z)
        - J_\nu (az)Y_\nu (z) \quad \beta =$
        twisted~, \\
        $ y_\nu (az)j_\nu (z) - j_\nu
            (az)y_\nu (z) \quad \beta =$ untwisted~.
    \end{minipage}
\right.
\eea
As in the RS model, the combinations of Bessel
functions relevant to the untwisted case are given by
\begin{eqnarray*}
    j_\nu (z) &=& {1\over 2}D(1-4\xi)J_\nu(z) +zJ_\nu'(z)~,\cr
    y_\nu (z) &=& {1\over 2}D(1-4\xi)Y_\nu(z) +zY_\nu'(z)~,
\end{eqnarray*}
This completes the Kaluza-Klein reduction of the bulk scalar
field.

In the following we will report only on the case of
untwisted fields, although the case of twisted fields can be
obtained at ease with simple modifications of our
calculation.

\section{Quantum Effective Action}
\label{qea}

The one loop effective action $\Gamma$ can be expressed as
the sum over the contributions of
each mode, $\Gamma_{\k,n}$
\bed
\Gamma = \sum \Gamma_{\k,n}~.
\eed
The previous expression can be evaluated in a variety of
ways (see for instance \cite{cw,djtplb83}).
Dimensional regularization of the $4-$dimensional Minkowski
directions to $4-2\epsilon$ leads to  the following expression
for the vacuum energy contribution to the
effective action 
\beq
\Gamma= - \int d^{4-2\epsilon}x \;V^{\rm reg}(s)~,
\label{eq15}
\eeq
with
\beq
  V^{\rm reg}(s) = -{1\over 2} (4 \pi)^{s} \mu^{2\epsilon}
  \Gamma(s)
  {\sum_{n,\k}}'  g_\k
  (q_{n}^2 +\lambda_\k^2)^{-s}~,
  \label{eq16}
\eeq
where the prime in the sum assumes that the zero mass
mode is excluded (since it does not contribute) and
$s=-2+\epsilon$.
The renormalization scale $\mu$ is
introduced for dimensional reasons.
It is convenient to separate $V^{\rm reg}$
into three contributions\footnote{Here we define $\lambda_0=0$, so that
  the existence of such a zero eigenvalue or not is
  controlled by $g_0$.  If $g_0\neq0$, the RS
  contribution comes about explicitly and introduces a
  divergence which needs to be cancelled by a corresponding
  contribution coming from $V_*$.  This cancellation
  provides a non-trivial check of our evaluation.  The case
  of a  strictly
  positive definite operator, can be obtained by putting
  $g_0$ to zero.}
\beq
V^{\rm reg}(s) =
V_\Sigma (s) + V_{RS}(s) + V_{*}(s) ~,
\label{splitting}
\eeq
where
\bea V_\Sigma (s) &=&- {\mu^{2\epsilon}\over 2(4\pi)^{-s}}
\Gamma(s) \sum_{\k=1}^{\infty} g_\k \lambda_\k^{-2s}~,\\
V_{RS}(s) &=& -g_0 {\mu^4\over 2(4\pi)^{-s}}
(ka/\mu)^{-2s} \Gamma(s) \sum_{n=1}^\infty x_n^{-2s}~,\\
V_{*}(s)&=& -{\mu^4 \over 2(4\pi)^{-s}}(ka/\mu)^{-2s} \Gamma(s)
\sum_{n,\k=1}^{\infty} g_\k \left( x_{n}^2 + y_\k^2 \right)^{-s} ~,
\eea with
$x_{n} = q_{n} / k a\,$ and $ y_\k = \lambda_\k / k a\,$.  Thus, $V_\Sigma$
retains the contributions from the orbifold zero mode (present only for an
untwisted field), $V_{RS}$ is the contribution from the $\Sigma$ zero mode
(which coincides with the potential in the RS model), and $V_*$ includes the
contribution from mixed states.  It is clear from Eqs.  (\ref{masses}) and
(\ref{fs}) that $x_n$ depends on $a$ only.  Since $\lambda_\k$ scales like
$1/R$, we can factor out the dependence on this modulus,
defining dimensionless eigenvalues $\hat \lambda_\k = R \lambda_\k$,
that do not depend on $R$. If $\Sigma$ is a one-parameter space,
then $\lambda_\k$ cannot depend on any other {\em shape} moduli. However, here
we are interested in the dependence on the {\em breathing} mode $R$ only.
So, in general, $y_\k=\hat \lambda_\k/(k a R)$ depends on the moduli described in
Sec. (\ref{moduli}) through $R_-$.

The first term in (\ref{splitting}) $V_\Sigma$ results from
the KK excitations along the internal manifold.  It can
be expressed in terms of the generalized $\zeta$ function
associated to the laplacian $P_\Sigma$ defined on $\Sigma$  (see Eq. (\ref{psigma})),
\footnote{The fact that this term does not depend on the
  eigenvalues $q_n$ is a consequence of how we have
  performed the Kaluza-Klein reduction.}
\beq
\label{zeta} \zeta(s) \equiv \zeta(s|P_\Sigma) =
\sum_{\k=1}^\infty g_\k \hat\lambda_\k^{-2 s} .
\eeq using the
previous rescaling we can recast $V_\Sigma$ as
\beq
V_\Sigma(-2+\epsilon) = -{1\over 32\pi^2 R^4} \left( \mu R
\right)^{2\epsilon} \Gamma(-2+\epsilon) \zeta(-2+\epsilon)~,
\eeq
where we redefined the renormalization constant $\mu$.
The previous expression can be elegantly dealt with by using the
Mittag-Leffler representation for the $\zeta$ function, which
proves to be a very useful tool to handle the pole structure of
the $\zeta$ function, since the residues at the poles are
determined by geometrical quantities of $\Sigma$ (See for example
\cite{blau}). As shown in appendix \ref{app:zeta},
\beq
 \zeta(s)
= {1 \over \Gamma(s)} \left\{  \sum_{p=0}^{\infty} {\tilde C_p
\over s - D_2/2 +p } + f(s) \right\}~,
\label{poleszeta}
\eeq
where  $\tilde C_p=C_p-g_0 \,\delta_{p, D_2/2}$ and $C_p$ are the
(integrated) Seeley-DeWitt coefficients of the operator $P_\Sigma$
on $\Sigma$, $p$ runs over the positive half integers and  $f(s)$
is an entire function. In fact, the sum (\ref{poleszeta}) runs
over half integers, but, since $\Sigma$ has no internal
boundaries, the coefficients $C_{i/2}$ are zero. Relation
(\ref{poleszeta}) can now be used to regulate $V_\Sigma$, and a
simple calculation gives
\beq
V_\Sigma (s=-2+\epsilon) = -{1 \over
32\pi^2 R^4} (4\pi R^2 \mu^2)^{\epsilon} \left[ \Omega_{-2} +
C_{D_2/2+2}{1\over \epsilon}\right],
\eeq
where $\Omega_{-2}$ is
the constant term in the power series of $\Gamma(s)\zeta(s)$
around $s=-2$ (see Eq. (\ref{omegap})).

The term proportional to the 
RS contribution has been computed in
\cite{flachi1,pujolas1,roth1}.
Without going into details, we write such term as follows:
\beq V_{RS}= -g_0  {k^4 \over 32\pi^2}
(k/\mu)^{-2\epsilon}
    \left\{-d_4 {1\over \epsilon}
        \left(
            1+a^{4-2 \epsilon}\right)
        + c_1 + a^{4} c_2 - 2 a^4 {\cal V}(a)
    \right\} ,
\label{rsone}
\eeq
where we have introduced
\beq
{\cal V}(a)= \int_0^\infty dz z^3
\ln\left(1-{k_\nu( z)\over k_\nu( a z)}
    {i_\nu( a  z)\over i_\nu( z)}\right)
\eeq
and  the coefficients $c_1$ and $c_2$ do not depend on $a$. Here, the
coefficient $d_4$ depends
on the mass and non-mininal coupling of $\Upsilon$, and is defined through
Eqs. (\ref{taylor},\ref{eq22},\ref{pdefinition}).

Before entering into the discussion of the higher dimensional
contribution due to the mixed KK states $V_*$, we can
foresee now some of the details of the computation.
As mentioned above,
the case when $\Sigma$ is a torus corresponds to a
toroidal compactification of a slice of higher dimensional $AdS$
space. Since it is a maximally symmetric space, all the geometric
invariants are constant, and so proportional to the brane tensions.
Thus, the only possible divergence that can appear is of the form
$\int d^4x (1+a^D)$. However, regardless of the dimension
of $\Sigma$, the contribution from $V_{RS}$ contains a divergence
of the form $\int d^4x (1 + a^4)$. Of course, what happens is that
aside from the higher dimensional divergence, $V_*$ also
contains another divergence that cancels the RS one.
This feature occurs not only when $\Sigma$ is a torus.
Rather, it is completely general. As we show next and
in appendix \ref{app:zeta},
the divergence of $V_*+V_{RS}$ proportional to
$\int d^4x (1 + a^4)$
is {\em always} controlled by a geometric invariant related to
$\Sigma$  (which trivially vanishes for a torus).
This ensures that if this divergence persists, it is because
one can build some operator that behaves like it in this background.


Let us now turn to the evaluation of $V_*$.
First of all, let us concentrate on the sum
\beq
\Gamma(s) \sum_{n,\k=1}^{\infty} g_\k
\left( x_{n}^2 + y_\k^2 \right)^{-s} ~.
\label{aux}
\eeq
This is not straightforward to compute, however
the method developed in \cite{lese,klaus1,klausbook} allows
us to perform such a calculation.  Since, in our case, the
evaluation does not present any particular difficulty, we
will be brief and address the reader to the original
references for an introduction to the details of the method.

The residue theorem permits us to express the sum
(\ref{aux}) as a contour integral and an appropriate choice
of the contour of integration leaves us with
\beq
\Gamma(s) {\sin(\pi s) \over \pi} \sum_\k g_\k
\int_{y_\k}^{\infty} (x^2 - y_\k^2)^{-s} {d \over dx}
\ln \left[F_\nu(ix)\right]dx ~,
\label{eq20}
\eeq
which, by changing variable and by using some known
properties of the
Bessel functions can be recast as
\beq
{1 \over \Gamma(1-s)} \sum_\k g_\k^{~}
y_\k^{-2s} \int_1^\infty (z^2 -1)^{-s} {d \over dz}
\ln \left[P_\nu(y_\k z)\right]dz ,
\label{eq21}
\eeq
where
\begin{equation}
    \label{pdefinition}
P_\nu(z)=F_\nu(i z)={2\over \pi}
\left[ k_\nu(z)i_\nu(a z)-k_\nu(a z)i_\nu(z)\right],
\end{equation}
and
\begin{eqnarray*}
    i_\nu(z)&=&zI'_\nu(z)+{1\over2}D(1-4\xi)I_\nu(z)\\
    k_\nu(z)&=&z K'_\nu(z)+{1\over2}D(1-4\xi)K_\nu(z).
\end{eqnarray*}
The integral (\ref{eq20}) is considered in \cite{contino} for $s=0$. In that case,
the result can be expressed in closed form. However, for $s=-2$ this does not
seem to be possible.

We can regulate relation (\ref{eq21}) using the
asymptotic  expansions for the Bessel functions. The large
$z$ behaviour of $i_\nu$ and $k_\nu$ can be written as
(see {\em e.g.}\cite{flachi1})
\bea
\label{besselasymptotics1}
i_\nu (z) &=&   \sqrt{z\over2\pi}e^z \Theta^{(i)}(z)~,\nn
k_\nu (z) &=& \sqrt{\pi z \over 2} e^{-z} \Theta^{(k)}(z)~,
\eea
where  $\Theta^{(k)}(z)=\Theta^{(i)}(-z)$
is a power series in $1/z$ beginning with $1$.
Thus, we can recast the integrand of Eq. (\ref{eq21}) in the form
\beq
  P_\nu (y_\k z) =- {\sqrt{a}\over \pi} y_\k z e^{(1-a)y_\k z}
  \Theta^{(i)}_{}( y_\k z)\Theta^{(k)}_{}(a y_\k z)
  \left[ 1 - {k_\nu(y_\k z) i_\nu(a y_\k z)
        \over i_\nu(y_\k z) k_\nu(a y_\k z)}
  \right]~.
  \label{eq22}
\eeq
Up to a constant term, $\ln P_\nu$ can be split as
$$
\ln\left[P_\nu (y_\k z) \right]=
{\mathcal H}_\k^{(1)} (z)+{\mathcal H}_\k^{(2)} (z)+{\mathcal
H}_\k^{(3)} (z)
$$
with
\begin{eqnarray}
    \label{eq23}
    {\mathcal H}_\k^{(1)} (z) &=&
    \ln z + (1-a) y_\k z ~,  \nonumber\\[1mm]
    {\mathcal H}_\k^{(2)} (z) &=&
    \ln\left[\Theta^{(i)}_{}( y_\k z)\Theta^{(k)}_{}
        (a y_\k z)\right] ~,    \nonumber\\[1mm]
    {\mathcal H}_\k^{(3)} (z) &=&
    \ln\left[ 1 - {k_\nu(y_\k z) i_\nu(a y_\k z)
          \over i_\nu(y_\k z) k_\nu(a y_\k z)}
    \right]~.
\end{eqnarray}
and correspondingly,
$$
V_*(s)=V_*^{(1)}(s)+V_*^{(2)}(s)+V_*^{(3)}(s) ,
$$
with
\begin{equation}
    \label{v*i}
    V_*^{(\alpha)}=
    -{\mu^4 \over 2(4\pi)^{-s}}(ka/\mu)^{-2s} {1\over \Gamma(1-s)}
    \sum_{\k=1}^{\infty} g_\k y_\k^{-2s}
    \int_1^\infty (z^2 -1)^{-s} {d \over dz}
    \ln \left[{\cal H}^{(\alpha)}_\k(z)\right]dz \qquad\quad (\alpha=1,2,3).
\end{equation}

The evaluation of $V_*^{(1)}(s)$ is analogous to the one for
$V_\Sigma(s)$ and, once more, the Mittag-Leffler
expansion allows to express the result in terms of the
heat-kernel coefficients of the operator $P_\Sigma$ on
$\Sigma$.
We find
\bea
V_*^{(1)}(s=-2+\epsilon)=
-{1 \over 32 \pi^2 R^4} \left(\mu R\right)^{2\epsilon}
&\Biggl\{&
\left[ {1\over2}C_{2+D_2/2} + {1\over 2\sqrt{\pi}}
C_{5/2+D_2/2}{1-a \over k a R } \right]
{1 \over \epsilon}  \nn
&+&
{1\over2} \Omega_{-2}
+{1\over2\sqrt{\pi}} \Omega_{-5/2}{1-a \over  k a R }
\Biggr\}
\label{e*1}
\eea

The second term $V_*^{(2)}(s)$ can be
evaluated\footnote{Strictly speaking we are using an
asymptotic expansion and therefore the equality sign is not
exact.
However, the approximation we are making is reasonable
because the integration range vary from $1$ to $\infty$ and
the argument of  $\Theta^{(i)}$ and $\Theta^{(k)}$
is large in the region $R \ll  1$ and $a R \ll 1$.}
using the explicit form of $\Theta^{(i)}$ and $\Theta^{(k)}$:
\beq
\ln \left( \Theta^{(i)}(z)\Theta^{(k)}(az) \right)
\simeq
\sum_{j=1}^{\infty}
\left(1+{(-1)^j\over a^j}\right)d_j~  \, z^{-j}\qquad
\text{for}\quad z\gg1,
\label{taylor}
\eeq
the coefficients $d_j$ can be obtained by simply Taylor
expanding the logarithm.
Using (\ref{taylor}) and treating
the sum over the eigenvalues $y_\k$ as in
the case of $V_\Sigma$ (see App. (\ref{app:zeta})),
we can write $V_*^{(2)}$ as
\beq
V_*^{(2)}(s=-2+\epsilon)=
{1 \over 32 \pi^2 R^4}  \left(\mu R\right)^{2\epsilon}
\sum_{j=1}^{\infty}
{d_j \over \Gamma(j/2)}
\left\{
\left[C_{2 + D_2/2 - j/2}- g_0\delta_{4,j}\right]
 {1 \over \epsilon}
 + \Omega_{-2+j/2}
\right\}
\Big((k a R)^j+(-k R)^j\Big) ~.
\label{eq27}
\eeq

The third term in $V_*^{(3)}(s)$ is finite by construction,
and we can  put safely $s=-2$,
\beq
V_*^{(3)}(s=-2)=-{1 \over 64 \pi^2 R^4} \sum_{\k=1}^\infty g_\k
\hat\lambda_\k^4
\int_1^\infty (z^2-1)^{2}{d\over dz}{\mathcal H}_\k^{(3)}(z)
dz~.
\label{eq28}
\eeq
Combining the previous results, we obtain the unrenormalized
Casimir energy:
\begin{eqnarray}
    \label{Ereg}
    V^{\rm reg}= -{1\over 32 \pi^2 R^4} &\Biggl[&
    \sum_{j=-1}^{\infty}
    \left[(k R_-)^j+(- k R_+)^j\right]
    \left\{ \gamma_j +(\beta_j -g_0 d_4 \delta_{4,j})
    {1\over \epsilon}
    \left(\mu R \right)^{2\epsilon} \right\}\nonumber \\
    &+& g_0 (k R)^4 \left\{ c_1 + a^{4} c_2 - 2 a^4 {\cal
V}(a)
        - {1\over \epsilon} \left(1+a^{4-2\epsilon}\right)
        \left(k/\mu\right)^{-2\epsilon}\right\}
    +2\sum_{\k=1}^\infty g_\k \hat\lambda_\k^4 {\cal
V}_\k(a,R_-)
    \Biggr]
\end{eqnarray}
where
\begin{equation}
    \label{betas}
    \beta_j=
    \left\{
        \begin{minipage}[c]{7cm}
            $(1/ 2\sqrt{\pi}) C_{5/2 + D_2/2} \quad$
            {\rm for}     \qquad  $j=-1$\\
            ${(3/ 2)} \ C_{2+D_2/2} \quad\qquad$ {\rm for}
            \qquad $j=0$\\
            $-(d_j / \Gamma(j/2))\,C_{2-j/2+D_2/2} \qquad
            \rm{otherwise}$,
    \end{minipage}
\right.
\end{equation}
and  we understand that the Seeley-DeWitt coefficients
$C_i$ are zero if $i<0$,
\begin{equation}
    \label{gammas}
    \gamma_j=
    \left\{
    \begin{minipage}[c]{7cm}
        $(1/ 2\sqrt{\pi}) {\Omega_{-5/2}} \quad$ {\rm for}
        \qquad  $j=-1$\\
        ${(3/ 2)} \Omega_{-2} \quad\qquad$ {\rm for} \qquad
        $j=0$\\
        ${-(d_j / \Gamma(j/2))}\Omega_{j/2-2} \qquad
        \rm{otherwise}$,
    \end{minipage}
\right.
\end{equation}
and
\begin{equation}
    \label{calvk}
    {\cal V}_\k(a,R_-)=\int_1^\infty dz ~z (z^2-1)
    \ln\left(1-{k_\nu(y_\k z)\over k_\nu(y_\k a z)}
        {i_\nu(y_\k a  z)\over i_\nu(y_\k z)}\right) .
\end{equation}
Equation (\ref{Ereg}) shows that as we advanced above,
the lower dimensional divergence coming from the
RS contribution $V_{RS}$ is always cancelled,
independently of the structure of the internal manifold
$\Sigma$. On the other hand, the contribution from the
KK modes along $\Sigma$ only (lower dimensional, as well)
may give a divergence corresponding to $j=0$. This is
controlled by the Seeley-DeWitt coefficient $C_{D_2/2+2}$,
and gives $1/2$ of the resulting $3/2$ factor in $\beta_0$,
the rest coming from the mixed states in $V_*$.
In particular, if $D_2$ is odd, then there is no such
divergence (if $\Sigma$ is boundaryless), in accordance
with the absence of any operator that behaves as
$\int d^4x \;1/R^4$ in the background, in this case.

To conclude this section, we briefly comment on the differences
appearing when we consider a twisted bulk field. First, since
there is no orbifold zero mode, its contribution $V_{\Sigma}$ is
not present. One can show that the asymptotic behaviour of the
function $P_\nu$ differs in two powers of the argument,
originating a change of sign in the contribution to $\beta_0$ and
$\gamma_0$ from $V_*^{(1)}$. Of course, the $d_j$ coefficients
also change, and can be read from \cite{flachi1,pujolas1}. In
brief, one needs to change the $d_j$ by the corresponding one, and
the $3/2$ factor in $\beta_0$ and $\gamma_0$ by $-1/2$. Also, we
haven't included any brane mass terms or kinetic terms, relevant for the untwisted
case only (aside from the ones arising from the coupling to curvature).
In principle, these can be different on each brane.
This changes our result in that we would have different
coefficients, $d_j^\pm$, for the $r_\pm$ series.

\section{Renormalization}
\label{ren}

In the previous section, we have computed the unrenormalized
Casimir energy (\ref{Ereg}) using dimensional regularization. This allows
us to isolate the divergent terms, of the form
\begin{equation}
    \label{sdiv}
    \Gamma^{\rm div}=
    {1\over\epsilon} {1 \over 32 \pi^2} {1\over R^4}
    \int d^4x \sum_{j=-1}^{D_2+4} \beta_j (a^j
    +(-1)^j)(kR)^j ,
\end{equation}
with $\beta_j$ given by (\ref{betas}). A finite number of
divergences appear because we have computed the one loop contribution to the
effective potential.

It is well known (see {\em e.g.} \cite{klausbook}) that the divergences present
in the effective action are given by the Seeley-DeWitt coefficient $C_{(D+1)/2}$
related to the operator in (\ref{eq3}) on our $D+1$ dimensional background
space-time. Since this has boundaries,  nonzero
boundary terms are present for any dimension.
Moreover, since the extrinsic curvature is constant in the space-time we are
considering, several powers of the intrinsic curvature of the boundaries are
present. Finally, it is easily shown that once any possible bulk term is
evaluated  on the background solution, it can be recast as boundary term for
this specific solution.\footnote{
  For instance, $\int d^{(D+1)}x \;\sqrt{g_{(D+1)}}\Lambda= \sum_\pm\int d^Dx
  \,\sqrt{g_{(D)\pm}} \,\sigma_\pm$, with $\sigma_\pm=\mp 2 \Lambda/ D k$.}

So, we shall consider  boundary term of the form
\beq
   \label{counttform}
   \sum_{\pm} \int d^{D}x \sqrt{g_{(D)\,\pm}}
   \,{\cal R}_{(D)\,\pm}^N ,
\eeq
where $N=0,1,2,\dots $  and  ${\cal R}_{(D)\,\pm}$ denotes the
(intrinsic) curvature computed from the induced metrics on
the branes $g_{(D)\pm}$.  Using the explicit expression for
the metric tensor (\ref{metricansatz}), a simple calculation
shows the previous term generates a contribution proportional
to $R^{D_2-2N}$ coming from the brane at $y_+$, and a
contribution of the form $R^{D_2-2N} a^{4+D_2-2\epsilon
  -2N}$ from the other brane.  Then, it is clear that all
the divergences in (\ref{sdiv}) can be dealt with operators of the
form (\ref{counttform}).
Specifically, we can take the following
expression as the counter-term needed to renormalize the
effective action:
\begin{eqnarray}
    S^{CT}_j &=&
    {1\over 32 \pi^2 \epsilon} \int d^{D_2}X
    d^{4-2\epsilon}x \left\{
        \sqrt{g_{(D)+}} \kappa^+_{j} {\cal R}_+^{(D_2+4-j)/2} +
        \sqrt{g_{(D)-}} \kappa^-_{j} {\cal
          R}_{-}^{(D_2+4-j)/2}
    \right\} \nonumber \\
    &=&{1\over 32 \pi^2 \epsilon}\int d^{4-2\epsilon}x
      {R^j\over R^4}
    \left\{\kappa^+_{j} + \kappa^-_{j}
        a^{j-2\epsilon}\right\}
    \label{counterterms}
\end{eqnarray}
The index $j$ here runs over the integers comprised between
$-1$ and $4+D_2$, and $\kappa^\pm_j$ are renormalization constants.
We recall that, from (\ref{sdiv}) and (\ref{betas}), the
divergences occur
for $j$ even only if  $D_2$ is even, and for $j$ odd when
$D_2$ odd.

In the process of subtracting the counter-terms, finite
contributions to the vacuum energy
with a logarithmic dependence on the moduli are generated.
The renormalized expression can be written as
\begin{eqnarray}
    \label{finite}
    V(R_\pm)= -{1\over 32 \pi^2 R^4} &\Biggl[&
    \sum_{j=-1}^{\infty} \Biggl\{
    \left( \beta_j-g_0\, d_4 \,\delta_{4,j}\right)
    \Big[(k R_-)^j\ln(k R_-)^2+(- k R_+)^j\ln(k R_+)^2\Big]
\nonumber \\
    &+& \left( \gamma_j - \beta_j \ln\left(k /\mu
\right)^2\right)
    \Big[(k R_-)^j+(- k R_+)^j\Big] \Biggr\} \nonumber \\
    &+& g_0 (k R)^4 \left\{ c_1 + a^{4} c_2 - 2 a^4 {\cal
V}(a) \right\}
    +2\sum_{\k=1}^\infty g_\k \hat\lambda_\k^4 {\cal
V}_\k(a,R_-)
    \Biggr]
\end{eqnarray}

A few remarks are now in order. First of all, note that we recast the result in
order to isolate the $\mu$ dependent terms. Such terms are not computable from
our effective theory, rather they have to be fixed by imposing a set of
renormalization conditions.  Secondly, notice that the result is valid for $D_2$
even as well as for $D_2$ odd, and the heat-kernel coefficients automatically
take this into account. Also, it should be noted that the flat space limit
$k\to0$ is well defined.  One has to take into account that $R_+= R e^{-k y_+}$
and a similar expression for $R_-$.  Then in the limit $k\to0$, keeping the
proper distance between two branes $(y_- - y_+)$ fixed, all terms which would be
singular in $k$ cancel each other in the first and second lines of Eq.
(\ref{finite}). The rest of the terms are well behaved.  The resulting
expression is proportional to $(y_- - y_+)/R^5$, which has a clear
interpretation in terms of the Casimir energy.

An important remark concerns the divergence proportional to $R_+^4+R_-^4$. This
is the divergence present in the RS contribution\cite{flachi1,pujolas1}.  For
$D_2$ odd, it is not reproduced by any of the counter-terms in
(\ref{counttform}).  However, this is not a problem because such divergence is
cancelled by the corresponding one coming from (\ref{eq27}) for $j=4$.

\section{Physical scales and the EW/Planck hierarchy}
\label{scales}

In this section we propose a scenario where supersymmetry is
broken at a scale $\eta_{\rm SUSY}$ not far below the cutoff scale
$M$, and the hierarchy between the electroweak and the effective
Planck scales is generated by a combination of redshift and large
volume effects. Also, we discuss the range of possible values for
the dynamical (the moduli $R_\pm$) and the fixed scales (the
cutoff $M$ and the SUSY breaking scale $\eta_{\rm SUSY}$).

From Eqs. (\ref{einsteinframe}) and (\ref{planckmass}), we see that the relation
between the four dimensional effective Planck mass and the higher dimensional
one (in the four dimensional effective theory using the Einstein frame metric
$\hat g_{\mu\nu}$) is
\begin{equation}
    \label{mplanck}
    m_P^2\approx 
    (M R)^{D_2}
    {M\over  k} M^2 .
\end{equation}
We shall assume that the masses of particles (located at $y=y_-$) are somewhat
below the cutoff $M$.
In the four dimensional theory, these masses
are redshifted down to $\sim a M$.
Then, the EW/Planck hierarchy is  given by
\begin{equation}
    \label{hierarchy}
    h^2\equiv a^2 {M^2 \over m_P^2}
    \sim {a^2\over (R M)^{D_2} }{k\over M}
    \sim 10^{-32}.
\end{equation}
Thus, the EW/Planck hierarchy $h$ is explained in this model due to a
combination of {\em redshift} \cite{RS1} and {\em large volume}
\cite{aadd} effects (even though the branes are of codimension 1).
The crucial ingredient in order for the large volume effect to be
efficient (aside from having a long orbifold), is that the additional
extra space $\Sigma$ exponentially grows as one moves away from the
negative tension brane (see Fig. 1).  In this way, matter is allowed
to propagate along a small $\Sigma$, of size $R_-$, whereas gravity is
diluted since it propagates through a much larger $\Sigma$, of
effective size $R_+$.
Since the gauge
interactions must not be diluted by an analogous effect, we have to
assume that the compactification scale on the negative tension brane
$1/R_-$ is close to the fundamental cutoff $M$.  \footnote{Keeping
  only the $\Sigma$-zero mode in the action for a $D$ dimensional
  Yang-Mills field $F_{AB}$ at $y=y_-$ with coupling constant
  $g^2_{{}_{*(D)}}\sim M^{4-D}$, one obtains $\displaystyle \int
  d^Dx\sqrt{g_{{}_{(D)-}}} \;{1\over g^2_{{}_{*(D)}}}
  F_{AB}F_{CD}\,g_{{}_{(D)-}}^{AC}\,g_{{}_{(D)-}}^{BD} \simeq \int d^4x
  \sqrt{\hat g} \;R_-^{D_2}\,{1\over g_{{}_{*(D)}}^2}\,
  F_{\mu\nu}F_{\rho\sigma} \hat g^{\mu\rho} \hat g^{\nu\sigma}$, where
  we used that $\tilde g_{\mu\nu}\simeq \hat g_{\mu\nu}$. Thus, the four
  dimensional YM coupling is identified as
  $g_{{}_{*(4)}}^2=g_{{}_{*(D)}}^2/R_-^{D_2}\sim 1/(M R_-)^{D_2}$.}

Our model solves  the hierarchy problem in a fashion very similar
to the models considered in \cite{chackoetal,chackonelson}, with
two concentric branes embedded in a noncompact bulk. In this
references, the hierarchy and the positions of the branes are
naturally stabilized by a generalization  of the
Goldberger and Wise mechanism \cite{gw} (see also
\cite{multamaki}).

\begin{figure}[htbp]
    \label{picture}
    \begin{center}
        \epsfysize=5 cm \epsfbox{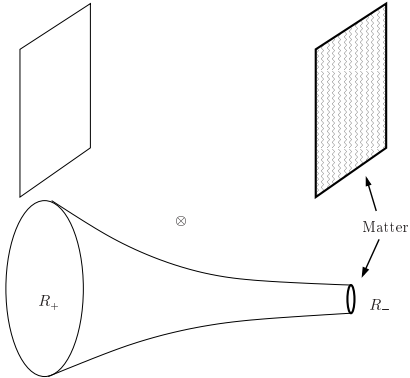}
        \caption{Matter can propagate
        along the additional extra space $\Sigma$ of size $R_-$,
        but gravity samples a much bigger space.}
    \end{center}
\end{figure}

Let us now examine the constraints that we have on the moduli and
the physical values they can take.  First of all, we are thinking
of an inter-brane distance $d=|y_--y_+|$ somewhat larger than the
inverse curvature scale $1/k$ of the bulk, in order to have a
substantial redshift factor $a=e^{-k d}$.  On the other hand, the
smallest physical length scale is given by the size of $\Sigma$ at
the negative tension brane, $R_-$. This cannot be smaller than the
fundamental length of the theory $M^{-1}$ though, as argued in the
previous paragraph, it should be close to it. There is a tighter
technical restriction which we shall use so that the result for
the potential (\ref{finite}) is valid. This, is organized as a
power series in $r_\pm=k R_\pm$, and can be trusted only when
$1/R_+$ a bit larger than the curvature scale and $1/R_-$ a factor
$a$ above (recall that the ratio $R_-/R_+$ coincides with the
redshift factor $a$).
Incidentally, we remark that this corresponds to the physical situation where
the size of the internal manifold $\Sigma$ is {\em everywhere} smaller than the
inter-brane distance $\sim 1/k$.  \footnote{This means that in a certain range
  of energy the model is effectively 5 dimensional.
  In appendix \ref{app:5dreduct}, we derive the form of the dimensionally
  reduced theory down to 5 dimensions.}
So, we must assume a separation between the fundamental cutoff
$M$ and the curvature scale $k$ at least of order $a$.  This leads to the
following scenario.

Consider a supersymmetric theory where the SUSY breaking
scale is given by $\eta_{\rm SUSY}$.
Then, the bulk cosmological constant $\Lambda \sim k^2 M^{D-1}$ is
expected to be proportional to $\eta_{\rm SUSY}^{D+1}$, which
leads to
\begin{equation}
    \label{keta}
    k\sim \left({\eta_{\rm SUSY}\over M}\right)^{(D+1)/2} M \ll M
\end{equation}
Even if SUSY is broken not far below the cut-off scale, this may
lead to a curvature scale $k$ many orders of magnitude below $M$,
due to the large exponent in (\ref{keta}). If the moduli $R_\pm$
are stabilized near the values $R_+\sim 1/k$ and $R_-\sim 1/M$,
then $a\sim k/M$ and from (\ref{hierarchy}), the hierarchy is
given by
\begin{equation}
    \label{ha}
    h \sim \left({k\over M}\right)^{(D-1)/2}\sim \left({\eta_{\rm SUSY}\over
    M}\right)^{(D^2-1)/4}.
\end{equation}
Note that the required hierarchy is obtained with $\eta_{\rm
SUSY}$ within one order of magnitude of the cut-off $M$ for
$D=10$, and less than 3 orders of magnitude below $M$ for $D=5$.

This shows how the problem of the stabilization of a large hierarchy works in
this model.  Having introduced a small separation between the SUSY breaking and
the cutoff scales, we obtain a stable very flat warped space-time, $k\ll M$.  If
the potential (\ref{finite}) can stabilize the moduli $R_\pm$ near the values,
$R_+\sim 1/k$ and $R_-\sim 1/M$, then the effective Planck mass is very
large as compared to the EW scale.
Whether or not  the effective potential (\ref{finite}) can do this job is
addressed in next section.

Let us discuss the physical scales in the model in some detail. As
illustrated in Fig. 1, the branes are of codimension 1, so that
matter (residing on the negative tension brane, at $y=y_-$) can
propagate through a physical extra dimensional space of size $\sim
R_-$. The mass scales on this brane are redshifted by a factor
$a$, thus the mass of the first KK excitations of matter fields is
$1/R$.  Then, from collider physics, we have to set the
compactification scale $1/R \gtrsim TeV$, at least.

In contrast, gravity propagates through the whole bulk space, and its KK
spectrum is analogous to that obtained in Sec. \ref{kkreduction} for a scalar
field. In particular, there are three kinds of modes, excited along the
orbifold only, along $\Sigma$ only or along both, as  (\ref{kkmasses}) shows.
The masses $m_{{}_{\Sigma}}$ of the first graviton KK modes along
$\Sigma$ are of order $1/R$.  However, the modes winding along the orbifold only
(the $\Sigma$ zero mode) have masses given by $m_{\rm orb} \sim a k$, as in the
RS model (the curvature scale times the redshift factor). In the approximation of everywhere
small $\Sigma$
that we are considering, $k R_\pm \ll 1$, this means that
these modes are a factor $a$ lighter than the modes propagating along $\Sigma$.

This allows us to assume the SUSY breaking scale $\eta_{\rm SUSY}$
and the cutoff $M$ are such that $k\sim TeV$, obtaining quite
small masses for the graviton orbifold KK modes $m_{\rm orb}\sim a
TeV$.  Below, we show that such a small value does not conflict
with observations, since the coupling of these modes to matter is
very suppressed.  This is consistent with the assumption made
above that in the higher dimensional theory the masses of matter
fields are near the cutoff $M$, since they are redshifted to $a M
\sim TeV$, which compatible with the electroweak scale.  Also,
since we are considering the limit of everywhere small internal
space $k R_\pm\lesssim 1$, setting $k\sim TeV$ implies that the
masses of matter KK fields is large enough, $1/R\gtrsim TeV$.

Thus, from the point of view of the
4 dimensional effective theory, KK modes from the matter fields appear at $1/R \sim
TeV$. 
Since the curvature scale $k$ of the bulk is close to $1/R$,
this coincides with the scale where gravity  becomes
higher dimensional.

In summary, we are lead to consider  distribution of scales  illustrated in Fig. 2.
We set the cutoff $M$ and the SUSY breaking scale $\eta_{\rm SUSY}\lesssim M$ such
that the curvature scale of the bulk is $k\sim TeV$.  We assume that some mechanism
can stabilize $R_-$ near the fundamental length $1/M$ and $R_+\sim 1/k$.
As a consequence, the masses of the
graviton KK modes along the orbifold
are $m_{\rm orb}\sim a TeV$, and for the modes along $\Sigma$ are
$m_{{}_{\Sigma}}\sim TeV$.
\footnote{Another interesting possibility consists of setting $m_{\rm orb} \sim
  TeV$ so that $k\sim a^{-1} TeV$ and $M\sim a^{-2} TeV$.  This could be realized
  in a scenario with the SUSY breaking scale $\eta_{\rm
    SUSY}\sim1/R\sim k$ and the masses of particles of order $k$, from the
  $(D+1)$ viewpoint.  In this scenario, the EW/Planck hierarchy is given by
  $h^2\sim a^{2} k^2/m_P^2$. If the moduli are stabilized so that $R_+\lesssim
  1/k$ and $R_-\gtrsim 1/M$, then $h \sim \left(k/M\right)^{D+1}$, thus needing
  less separation between $\eta_{\rm SUSY}$ and $M$ in order to explain same
  hierarchy $h$. Moreover, one can see that the potential (\ref{finite})
  generates masses for the moduli larger than in the scenario presented so far.
  However, the bulk cosmological constant $\Lambda$ would be much larger than
  $\eta_{\rm SUSY}^{D+1}$.}

\begin{figure}[htbp]
    \label{figscales}
    \begin{center}
        \psfrag{rmin}{$1/R_-$}
        \psfrag{E}{$E$}
        \psfrag{M}{$M \gtrsim  \eta_{\rm SUSY} \sim a^{-1} TeV$}
        \psfrag{mS}{$m_{{}_\Sigma}\sim 1/R$}
        \psfrag{mRS}{$m_{\rm orb}\sim a k\sim a TeV$}
        \psfrag{k}{$k\sim  TeV$}
        \psfrag{a}{$a$}
        \epsfysize=5 cm \epsfbox{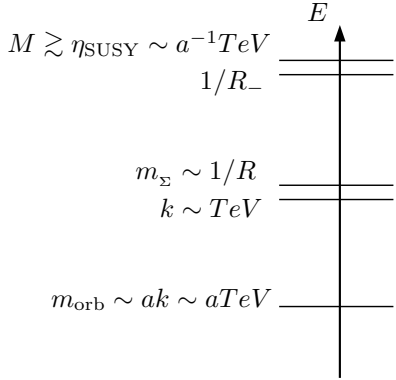}
        \caption{The mass $m_{\rm orb}$ of the first KK
          excitations along the orbifold is much smaller
          than the mass $m_{{}_\Sigma}$ of the modes excited
          along the internal manifold. There is essentially
          the same hierarchy between the fundamental cutoff
          $M$ and the scales that determine the solution,
          $k$ and $R$.}
    \end{center}
\end{figure}


It remains to be seen that, indeed, the graviton KK modes along the orbifold are
unobservable, in spite of their relatively small masses $m_{\rm orb}\sim a k$.
We see from Eqs.  (\ref{eq11},\ref{eq12}) that the (unnormalized) wave function
of the KK modes grows exponentially as $e^{D k |y| /2}$, signaling that the more
warped the extra dimensions are, the more localized on the negative tension
brane these modes are.  This implies \cite{rizzo} that the coupling of these
graviton KK modes is amplified with respect to that of the zero mode ($\sim
1/m_P$) by a factor $a^{-D/4}=h^{-D/2(D-1)}$. Then, they are much more weakly
coupled to matter $\sim 1/(10^{8}TeV)$ than in the RS model ($\sim 1/TeV$).
Thus, in spite of their relatively small mass, these KK gravitons cannot be seen
individually in accelerators. 
Moreover, since they are associated with only one off-the-brane dimension,
they do not have as large a multiplicity as in the usual large volume mechanism
\cite{aadd}, and so they do not significantly cool stars.  The total rate of
emission of any of such gravitons at a given energy $E<TeV$ can be estimated as
the coupling squared times the number of states with masses lighter than $E$
\cite{aadd,addphen},
$$
\left({1\over a^{D/4} m_P}\right)^2 {E \over m_{\rm orb}}
\sim h^{(D-4)/(D-1)} {E\over TeV} {1\over TeV^2},
$$
which is very small for the  energies available  inside stars.



\section{Stabilization}
\label{stab}

The result we have obtained so
far~(\ref{finite})
is a potential $V(R_\pm)$ for the two moduli describing the
background.
Using $r_\pm\equiv k R_\pm$, it can be cast as
\beq
\label{vRr}
V(r_\pm)=-{1\over 32\pi^2 R^4} \left[ V_+(r_+)+ V_-(r_-)+
v(r_+,r_-)\right],
\eeq
where $v(R,r)$ contains the 'non-local' part, and
\beq
\label{vRpm}
V_\pm(r_\pm)=\sum_{j=-1}^\infty (\mp 1)^j \left\{ \gamma_j
r_\pm^j + \left(\beta_j  -g_0 d_4 \delta_{j,4} \right)
r_\pm^j \ln{r_\pm^2} - \alpha_j^\pm r_\pm^j \right\},
\eeq
Here, the coefficients $\alpha_j^{\pm}$ are understood to be finite
renormalization constants,
and are nonzero when the corresponding logarithmic term is nonzero.
This is dictated by $\beta_j$ being zero or not ({\em i.e.}, whether
or not such a term is divergent), with the
sole exception of $j=4$. If  $\beta_4=0$ and
the laplacian  $P_\Sigma$ (see Eq. (\ref{psigma})) has one zero
eigenvalue, $g_0=1$, the logarithmic terms corresponding to
$j=4$ are not associated to any divergence of the effective
action, and $\alpha_4^{\pm}=0$. This situation arises, for example,
when $\Sigma$ is a torus.

Note that the sum goes from $-1$ to $\infty$ and we recall that from
Eq. (\ref{betas}), all the $\beta_j$ with $j>4+D_2$ vanish identically.
Thus, the
term $\beta_j r_-^j$ appears with $j$ running from $-1$ to
$D_2+4$, and the
same holds for the terms with $\alpha_j^{\pm}$ (there are
a finite number of divergent terms).

One interesting feature of the effective potential
(\ref{vRr}) in both cases
with $D_2$ even and odd is that the two
leading terms in the small  $r_\pm$ limit (corresponding
to $j=-1,0$) do not
depend on the mass $m$ nor the non-minimal coupling constant
$\xi$. This means that if
we consider  equal number of fermionic and bosonic
degrees of freedom, these terms cancel identically
even with non supersymmetric masses.
From now on, we will
focus on this case, one motivation being that the models
considered here arise
mainly in string theories, and the field content of the
effective theories
indeed contain equal number of bosonic and fermionic degrees
of freedom.  The
only change is that the sum in Eq. (\ref{vRpm}) will begin at
$j=1$ instead of $j=-1$.
As mentioned above,  the effective potential contains a
finite number of renormalization parameters
$\alpha^{\pm}_j$.
Their values are not computable from our effective theory.
Rather, we shall fix them
by requiring some renormalization conditions,
which determine the values for the moduli as well.
Since the moduli must be stabilized, we demand
\beq
\partial_{r_+} V(r_\pm)= \partial_{r_-} V(r_\pm)=0~,
\label{RCmin}
\eeq
and in order to match the observed value of the
effective four dimensional cosmological
constant, we shall impose
\beq
\label{RCzeropot}
V(r_\pm)|_{\rm min}\simeq 10^{-122} m_P^4.
\eeq
We are interested in the limit when the size of $\Sigma$ is
everywhere smaller than
the orbifold size, $r_+\lesssim 1$ and $r_-\ll 1$.
One can show
\footnote{
For instance, consider a the six dimensional example,
with $\Sigma=S^1$. As described in more detail in Subsect.
\ref{flatsigma}, the generalized zeta function is related to
the Riemann zeta function.
In this case the
we can easily work out the asymptotic behaviour of the
nonlocal contribution due to  the mixed KK states
${\cal   V}_{\k}(a,R_-)$ defined in (\ref{calvk}).
If we keep the first term in the
asymptotic expansion of the Bessel functions
(\ref{besselasymptotics1})
\begin{eqnarray*}
    {\cal V}_\k(a,R_-) & \sim & \int_{1}^\infty  dz
    (z^3-z)
    \ln\left( 1 - e^{-2(1-a)y_\k z}\right) \cr
    &=&-{1\over8}{ 1\over y_\k^4 (1-a)^4} \left\{4\
    (1-a)^2 y_\k^2 {\rm Li}_3
        \left( e^{-2 (1-a)y_\k}\right) +
        6\ (1-a) y_\k\ {\rm Li}_4\left( e^{-2
    (1-a)y_\k}\right) +
        3\ {\rm Li}_5\left( e^{-2 (1-a)y_\k}\right)
    \right\}.
\end{eqnarray*}
Taking only the first term in the series of the poly-logarithms for small
arguments ${\rm Li}_n(z)\approx z$, recalling that $y_\k=\hat\lambda_\k/r_-$
with $\hat\lambda_\k=\k$, $g_\k=2$ and summing over $\k=1,2,\dots$, we find to
leading order
\begin{equation*}
     \sum_{\k=1}^\infty
    g_\k \hat \lambda_\k^4  {\cal V}_\k(a,R_-)
    \sim
    -{1\over (1-a)^2}{r_-^2}
    e^{-2(1-a)/r_-}.
\end{equation*}
Thus, this contribution is safely negligible in the limit of
small internal space size $r_-\ll 1$.
}
that in this limit the non-local term
$v(r_\pm)$ is exponentially
suppressed, and we can approximate  the potential by the
'local' terms $V_\pm(r_\pm)$.
Moreover, since we consider only the positive powers of
$r_\pm$ in $V_\pm$, the potential at the minimum is
dominated by $r_+$.
Then, conditions (\ref{RCmin}) and (\ref{RCzeropot}) reduce
to
\begin{equation}
    \label{RCapprox}
    V_+'(r_+)= V_-'(r_-)=0,
    \qquad\text{and}\qquad
    {R^{-4}} V_+(r_+)|_{\rm min} \simeq 10^{-122} m_P^4.
\end{equation}
To investigate whether this potential can stabilize the moduli, we
consider separately the  cases with flat and curved $\Sigma$.


\subsection{Flat $\Sigma$}
\label{flatsigma}



This case corresponds to a toroidal compactification of a $4 + D_2 +1 $
dimensional RS model (with two codimension one branes).
In this case, all the divergences have the same form, because all
geometric invariants are constant and thus proportional to the brane tensions.
Thus, there will appear a logarithmic term in the $(4+D_2)$-th power of
$r_\pm$.
As  can also be derived from Eqns. (\ref{finite}),
(\ref{betas}) and (\ref{gammas}), setting $C_j=0$ for all $j\neq0$ ,
and $g_0=1$, there is another logarithmic term corresponding to $j=4$.

Thus, the expression for the
potential reduces to

\begin{eqnarray}
    \label{vRTorusII}
    V_{\pm}(r_\pm)&\approx&
    \Big\{ \mp\gamma_1 r_\pm + \gamma_2 r_\pm^2
    \mp \gamma_3 r_\pm^3
    + \left(\gamma_4 -  d_4  \ln{r_\pm^2} \right) r_\pm^4
    +\dots\nn
    &+& (\mp1)^{4+D_2} \beta_{4+D_2} r_\pm^{4+D_2}
    \ln{r_\pm^2}
    - \alpha_{4+D_2}^\pm r_\pm^{4+D_2} +\dots \Big\}.
\end{eqnarray}
To illustrate better how the stabilization mechanism works
in these cases, we shall
discuss in more detail the six
dimensional example with $\Sigma=S^1$.

The laplacian (\ref{psigma}) on this flat manifold is
$P_\Sigma=\partial_\theta^2/R^2$, and its generalized zeta
function (\ref{zeta},\ref{a1}) is related to the Riemann
zeta function through
$$
\zeta(s|\partial_\theta^2)=2\zeta_R(2s).
$$
The pole structure of $\zeta_R(2s)$ is easily found and
one immediately identifies
\begin{equation}
    \label{betascircle}
    \beta_j=
    \left\{
        \begin{minipage}[c]{7cm}
            \begin{tabbing}
            $-4 d_5 / 3 $  \=\qquad\=\qquad\=\qquad
            {\rm for}       $j=5$,\\
            0 \>\qquad\>\qquad\>\qquad
            \rm{otherwise},
            \end{tabbing}
    \end{minipage}
\right.
\end{equation}
and
\begin{equation}
    \label{gammascircle}
    \gamma_j=
    \left\{
    \begin{minipage}[c]{9cm}
        \begin{tabbing}
            ${-2/945}$   \=\qquad\=\qquad\=\qquad
            {\rm for}          $j=-1$,\\
            $3 \zeta_R'(-4)$ \>\qquad\>\qquad\>\qquad
            {\rm for}                  $j=0$,\\
            $4\zeta_R'(-2)d_2$ \>\qquad\>\qquad\>\qquad
            {\rm for}          $j=2$,\\
            $-d_4 $ \>\qquad\>\qquad\>\qquad
            {\rm for}
              $j=4$,\\
            $\displaystyle -{8 d_j \zeta_R(j-4) \over (j-4)(j-2)}$
            \>\qquad\>\qquad\>\qquad \rm{otherwise}.
        \end{tabbing}
    \end{minipage}
\right.
\end{equation}

From (\ref{vRTorusII}), the potential is of the form
\begin{equation}
    \label{vRcircle}
    V_{\pm}(r_\pm)\approx
    \Big\{ \mp\gamma_1 r_\pm + \gamma_2 r_\pm^2
    \mp \gamma_3 r_\pm^3
    + \left(\gamma_4 -  d_4  \ln{r_\pm^2} \right) r_\pm^4
    \mp \beta_{5} r_\pm^{5}
    \ln{r_\pm^2}
    - \alpha_{5}^\pm r_\pm^{5} +\dots \Big\}.
\end{equation}
As we mentioned above, the renormalization constants
$\alpha_5^\pm$ arise from a finite renormalization
$\delta\tau_\pm$ of the
brane tensions,
$$
\delta\tau_\pm \int d^5x \sqrt{g_{(5)\pm}}\,
={2\pi\over R^4}\int d^4x \,\delta\tau_\pm R_\pm^5,
$$
so that $\alpha_5^\pm=2\pi \delta\tau_\pm/k^5$.  The size of $\delta\tau_\pm$
is expected to be set by the SUSY breaking scale $\eta_{\rm SUSY}$
so that $\alpha_5^\pm$ are large in principle.  Then, the main contributions to
this potential arise from the fifth and the first powers.  The extremum
condition for the $r_-$ modulus can be well approximated by
$$
{\delta\tau_-}\simeq {\gamma_1\over10\pi}
{1\over r_-^4} k^5.
$$
Setting the natural value $\delta\tau_-\sim \eta_{\rm SUSY}^{5}$, we obtain
$$
r_-\sim \left({M\over \eta_{\rm SUSY}}\right)^{1/2}{k\over M},
$$
so that indeed $R_-$ is stabilized just above the fundamental scale $1/M$ without fine
tuning.

As for $r_+$, we have two conditions for just one variable, $\delta\tau_+$.  The
idea is to use the renormalization constant $\delta\tau_+$ in order to satisfy
$V_+|_{\rm min}\simeq 0$, and then using this value in $V_+'=0$, the $r_+$ is
determined. In order to be consistent, we should obtain $r_+\lesssim 1$. In such
a case, we can foresee from Eq. (\ref{vRcircle}) that if $\alpha_5^+$ has to
compensate for the potential at the minimum, it has to be of order one. But this
means that $\delta\tau_+$ is fine tuned to a value $\sim k^5$ instead of
$\eta_{\rm SUSY}^5$.

Imposing explicitly these conditions, we obtain
$$
{\delta\tau_+}\sim{\gamma_1\over 6\pi}{1\over r_+^4} k^5\sim k^5,
$$
and
$$
    r_+ \simeq  {4\gamma_1\over 3\gamma_2}\sim1.
$$
We can easily check that for a twisted field this ratio is $\simeq 0.6$, in
agreement with the assumption we made above.  For the untwisted case, this ratio
depends on the boundary and bulk masses, so it can be made small generically.
In conclusion, besides the  fine tuning  needed in order to match the four
dimensional cosmological constant, no tuning is needed for the  Planck/EW
hierarchy in this case.


A simple computation gives the mass that this potential induces for the canonical
moduli $\Phi$ and $\Psi$ of section \ref{moduli}
\begin{eqnarray}
   \label{modulimasses}
   m_\Phi^2 &\simeq& -{\gamma_1 \over 24 \pi^2 }
    {k \over  R^{3} m_P^2}
   \sim -\gamma_1 (h k)^2 \sim (1/mm)^2\\
   m_\Psi^2 &\simeq& -{ \gamma_1 \over 192
     \pi^2\,}
   \,a^{-2} \, {k\over  R^{3} m_P^2  }
   \sim -\gamma_1  (a TeV)^2,
\end{eqnarray}
where we used $1/R\sim k\sim TeV$ and Eq. (\ref{ha}).  The mass of
$\Phi$ is of the order of the inverse
millimeter, which is
large enough in order not to cause deviations from
Newton's law at short distances. 
Since, as shown in section \ref{moduli},
the coupling of $\Phi$ to matter is suppressed by
a Planckian factor, its effects in accelerators are
negligible as well.
On the other hand, the mass for the modulus $\Psi$ is of $10 KeV$ size. From
(\ref{eommoduli}), its coupling to matter is suppressed as $1/(10^4 TeV)$.

Let us briefly discuss the stabilization when we consider a higher dimensional
flat $\Sigma$.
As mentioned above, with more flat dimensions,
the renormalization constants related to the brane tensions
$\alpha_{D}^\pm$ appear with  higher powers of $r_\pm$.
The only change with respect to the case above is that the condition $V_-'=0$
now reads
$$
\delta\tau_-\sim{1\over r_-^{D-1}} k^D,
$$
and assuming a natural value for $\delta\tau_-$ given by $\eta_{\rm
  SUSY}^{D}$, we obtain again $r_-\sim (k/M)(M/\eta_{\rm SUSY})^{1/2}$. Thus,
for any dimension $D$ the modulus $R_-$ is stabilized without fine tuning near
$1/M$.

As for the modulus $R_+$, we expect the potential (\ref{vRTorusII}) to stabilize
it near $k$ once the fine tuning of $\delta\tau_+\sim k^D$ needed for the cosmological
constant is performed.

We can compute the masses for the moduli for an arbitrary number of flat
internal dimensions. We find that the mass for the $\Phi$ is always millimetric,
whereas $m_\Psi\sim a TeV$ increases with $D_2$, ranging from $10 KeV$ for
$D_2=1$ to $100 MeV$ for $D_2=6$.  The coupling of $\Psi$ to matter, of strength
(see Eq. (\ref{eommoduli}))
$$
1/\left( h^{-1/(D-1)} \;TeV \right)~,
$$
is comprised between $ \sim 1/(10^{4} TeV)$ for $D_2=1$
and $\sim 1/( 100 TeV)$ for large $D_2$.
This guarantees that it hasn't been produced at
colliders, or has any effect in star cooling.

\subsection{Curved $\Sigma$}
\label{curvedsigma}

When $\Sigma$ is not flat,  besides the
divergences proportional to brane tensions terms (giving rise to the power
$r_\pm^D$ in $V_\pm$), the potential has more divergences. For instance,
there can appear divergences proportional to curvature terms, which  give rise to the powers
$r_\pm^{D-2}$. Accordingly, terms with fewer powers of $r_\pm$ are due to higher
powers of the curvature, and in general the effective potential takes the form
\beq
\label{vcurvedsigma}
V_\pm(r_\pm)=\sum_{j=1}^\infty (\mp 1)^j \left\{ \gamma_j
r_\pm^j + \left(\beta_j r_\pm^j -g_0 d_4 \delta_{j,4} \right)
\ln{r_\pm^2}  \right\}  - \sum_{j=1}^D \alpha_j^\pm r_\pm^j.
\eeq

As in the previous case for the brane tensions, the size of the renormalization constants in
front of these operators are expected to be of order the cutoff scale $M$ (or $\eta_{\rm SUSY}$). Finite
renormalization terms of boundary operators behave as,
$$
M^{j}
\int d^Dx \sqrt{g_{(D)\pm}} \,{\cal R}^{(D-j)/2}\,
={1\over R^4}\int d^4x \,( M R_\pm)^{j}
={1\over R^4}\int d^4x \, \alpha^\pm_j r_\pm^{j},
$$
and we conclude that the dimensionless renormalization constants in
(\ref{vcurvedsigma}) are large,
$\alpha_j^\pm\sim (M/k)^j\gg1$.
Thus, these
terms are a series in
$M  R_\pm > 1$
rather than in $k R_\pm < 1$,
the dominant terms being with the highest powers, that is, the
brane tension and the curvature terms.
As a first approximation, we can neglect the remaining terms, and minimum
condition for $R_-$ is reached naturally for $R_-\sim 1/M$, which is what we
need (see Fig. 2).

However, we see that in order to obtain $R_+\sim 1/k$, we need to tune the ratio
of $\alpha_D$ and $\alpha_{D-2}$. Besides, the tuning corresponding to the
cosmological constant is still needed.

In principle, we could consider the case when the heat kernel coefficient
$C_{1}(P_\Sigma)$ is zero, which can happen for some value of the non-minimal
coupling $\xi$. We see from (\ref{betas}) that in this case there is no
divergence in the potential corresponding to the curvature terms.\footnote{The
  same thing cannot happen for the brane tension terms, since the corresponding
  coefficient is $C_0(P_\Sigma)=1$ always.}  Then, assuming that the next
nonzero coefficient is $C_2$, the two powers that  dominate the potential
are $(MR_\pm)^D$ and $(MR_\pm)^{D-4}$.  However, in order to stabilize $R_+$
near $1/k$, again we have to do one fine tuning.  We can say that in general,
the presence of any other divergence, besides the brane tension, spoils the
efficiency of the potential in stabilizing the moduli at well separated scales.

We conclude that, for curved $\Sigma$
the potential can naturally stabilize the moduli but without a large hierarchy.





\section{Conclusions}

In this article we have investigated the role of quantum effects
arising from bulk scalar fields in higher dimensional brane
models. Specifically, we have considered a class of warped brane
models whose topology is ${\cal M} \times \Sigma \times S^1/Z_2$,
where $\Sigma$ is a $D_2$ dimensional one-parameter compact
manifold, ${\cal M}$ is the four dimensional Minkowski space and
both ${\cal M}$ and $\Sigma$ directions are warped as in the
Randall Sundrum model, with two branes of codimension one sitting
at the orbifold fixed points. Aside from the usual negative
cosmological constant, a bulk sigma model scalar field theory is
used as the source of gravity in the cases of a curved internal
manifold $\Sigma$. We have identified the relevant moduli fields
characterizing the background, and found the classical action in
the moduli approximation (as well as the coupling of the moduli to
matter fields sitting on the branes).

We have computed the contribution to the one-loop effective action
from generic bulk scalar fields at lowest order (i.e. the Casimir
energy). The computation, similar to the one for the RS model, is
technically more complex since there are KK modes propagating
along $\Sigma$, resulting in a dependence of the Kaluza-Klein
masses on the eigenvalues of the Klein-Gordon operator on
$\Sigma$. However, for the specific choice of space-time we made,
where the warp factors for $\Sigma$ and for the Minkowski factor
${\cal M}$ are the same, the physical KK masses split as in the
usual factorisable geometries.  Using the Mittag-Leffler expansion
for the generalized $\zeta-$function we were able to express the
Casimir energy in terms of heat-kernel coefficients of the
internal space $\Sigma$, so that the presence of each divergence
is dictated by a certain heat-kernel coefficient. This simplifies
the renormalization of the result. An interesting nontrivial check
of our result is the fact that the RS divergence (which is lower
dimensional in this model and which appears as the contribution of
the $\Sigma$ zero mode) cancels out in the final result, as it
should, once all contributions are added. We renormalized the
effective potential by subtracting suitable counter-terms
proportional to a number of boundary or bulk local operators.
Since we work in dimensional regularization, the subtraction is
performed in the regularized space, with $(4-2\epsilon)+D_2+1$
dimensions.  As a result, there is a mismatch in the powers of the
moduli appearing in the divergent terms, and a number of
logarithmic terms (in the moduli) appear in the renormalized
expression for the effective potential.

As an application, we proposed a scenario where SUSY is broken at
a scale just below the fundamental cutoff $M$. This makes the
curvature scale of the background to be several orders of
magnitude below $M$. As a result, a large hierarchy is generated
by a combination of redshift \cite{RS1} and a large volume effects
\cite{aadd}. The key point for the latter to be efficient (in
spite of having codimension one branes) is that the size of the
internal manifold $\Sigma$ (present in the bulk and on the brane)
grows as one moves away from the TeV brane, where matter lives.
Therefore, this behaves effectively as a brane with a {\em small}
$\Sigma$ extra space, attached to which there is a {\em large}
$\Sigma$ space where only gravity propagates.

As for the stabilization, we find that, generically, the potential induced by
bulk fields can generate sizeable masses for the moduli compatible with a large
hierarchy with no need of fine tuning if $\Sigma$ is flat. If it is curved, the
effective potential can naturally stabilize the moduli but without a large
hierarchy.

In the model we have considered, the size of the internal space
$\Sigma$ is everywhere small compared with the size of the
orbifold. Therefore, there is a range of energy scales where the
model is effectively five dimensional (this feature is in common
with the Ho\v rava-Witten model \cite{hw,stelle}). From the
five-dimensional point of view, the model contains a dilaton field
in the bulk, which causes a power-law warp factor in the Einstein
frame, $a(y)\propto y^q$, where $y$ is the proper distance along
the extra dimension. The power $q$ is related to the number of
additional dimensions through $q=(D_2+3)/D_2$ \cite{pujolas2},
which leads to $1<q\leq 4$. Five dimensional models with power-law
warp factors were investigated in \cite{pujolas2}, where it was
argued that the counterterms at the orbifold fixed points can
naturally stabilize the moduli corresponding to the positions of
the branes. However, a large hierarchy was not expected unless the
power is substantially large, $q\gtrsim 10$. This conclusion is
consistent with the results of the present paper, which correspond
to relatively small $q$. In this case, the large hierarchy can
only be stabilized naturally if the internal space is flat. This
case is special because the only possible counterterms are
renormalizations of the higher dimensional brane tensions.

The above arguments suggest that a large hierarchy may be obtained
by considering more general warped models, where a larger power
exponent $q$ is obtained after reducing to five dimensions. In
such cases, the stabilization of a hierarchy without fine tuning
is expected even if the internal manifold $\Sigma$ is curved and
all sorts of counterterms are present. As a continuation of the
work presented here, we are currently considering the case in
which the warp factor along $\Sigma$ is constant
\cite{gregory,daemi}. We hope to report on this shortly
\cite{forthcoming}.

\acknowledgements

J.G. and O.P. are grateful to Alex Pomarol for very useful
discussions. A.F. acknowledges discussions with Y. Himemoto, M.
Sasaki and W. Naylor and the members of the Department of Earth
and Space Science at Osaka University for the kind hospitality.
A.F. is supported by the European Community via the award of a
Marie Curie Fellowship. We acknowledge support by the DURSI
Research
Project 2001SGR00188,  by CICYT under Research Projects AEN99-0766, and the MCyT
and FEDER under projects
FPA 2002-3598, FAP 2002-00748 and FAP 2002-00648, and HPRN-CT-2000-00131.
T.T. is supported by a Monbukagakusho Grant-in-Aid No.~14740165.
To complete this work, the discussion during and after the
YITP workshop YIYP-W-01-15 on ``Braneworld - Dynamics of
spacetime boundary'' was useful.

\appendix

\section{Mittag-Leffler expansion}
\label{app:zeta}

In the present appendix we prove Eq.
(\ref{poleszeta}).
It is well known that for a
strictly positive definite laplacian $P$ with eigenvalues
$\lambda_P$, the associated
zeta function $\zeta(s|P)=\sum \lambda_P^{-s}$
admits a  {\em Mittag-Leffler} expansion of the form
\beq
\zeta(s|P) = {1\over \Gamma(s)} \left\{ \sum_{p=0}^\infty
{C_p(P)\over s-D_2/2 +p} + f(s|P)\right\}
\label{mtgeneral}
\eeq
where $C_p(P)$ are the Seeley-DeWitt coefficients
related to $P$, with $p\in{\mathbb N}/2$
and the function $f(s|P)$ is analytic for all finite $s$.
This is a very useful relation since it neatly shows the
pole structure of the zeta function, in terms of geometrical
invariants.

We need to generalize this equation to operators with one
zero eigenvalue ($g_0=1$).
Consider a positive semidefinite differential operator
$P_\Sigma$ with eigenvalues $\lambda_\k^{2}$ and
assume that there is one zero eigenvalue.
As usual in these cases, one defines the
generalized
$\zeta$ function excluding this eigenvalue (see Eq.  (\ref{zeta})),
\beq
\zeta(s) = \sum_{\k=1}^\infty \hat\lambda_\k^{-2s} .
\label{a1}
\eeq
Our main task is to express  $\zeta(s)$ in
terms of geometrical quantities in the form
\beq
\zeta(s) = {1\over \Gamma(s)} \left\{ \sum_{p=0}^\infty
{\tilde{C}_p \over s-D_2/2 +p} + f(s)\right\} ,
\label{a2}
\eeq
with $p\in{\mathbb N}/2$, for some $\tilde{C}_p$ related
to the Seeley-DeWitt coefficients of $P_\Sigma$.
First of all, let us introduce the regulated
zeta function associated to  the operator
\footnote{Note that this mass  is fictitious and has nothing
  to neither with the physical bulk mass
  $m$ (\ref{eq3}) nor with the renormalization scale.}
$P_\Sigma^\mu\equiv P_\Sigma+\mu^2/R^2$,
\beq
\zeta_\mu(s) = \sum_{\k=0}^\infty (\hat\lambda_\k^{2} +
\mu^2)^{-s} .
\label{a3}
\eeq
Now it is trivial to express the function  $\zeta(s)$  in
terms of $\zeta_\mu(s)$,
\beq
\zeta(s) = \lim_{\mu \rightarrow 0} \left(\zeta_\mu(s) -
\mu^{-2s} \right) ,
\label{a4}
\eeq
and it is obvious that,  understood
as this limit, $\zeta(s)$
is infrared finite for any $s$, even though
$\zeta_\mu(s)$ is only IR finite for ${\rm Re}(s)\leq0 $.
By construction, $P_\Sigma^\mu$ is strictly positive
definite, so
$\zeta_\mu(s)$ admits the expansion
\beq
\zeta_\mu(s) = {1\over \Gamma(s)} \left\{ \sum_{p=0}^\infty
{C_p(\mu)\over s-D_2/2 +p} + g(\mu,s)\right\},
\label{mtmassive}
\eeq
where the function $g(\mu,s)$ is analytic
and the Seeley-DeWitt coefficients $C_p(\mu)$
depend polynomially on $\mu$.

From Eqns.(\ref{a4},\ref{mtmassive})
it follows that  \beq
\zeta(s) = {1\over \Gamma(s)} \lim_{\mu \rightarrow 0}
\left\{ \sum_{p=0}^\infty {C_p(\mu)\over s-D_2/2 +p} +
g(\mu,s)
- \Gamma(s) \mu^{-2s} \right\}.
\label{a5}
\eeq
The next step is to isolate the poles from the last term in the
previous formula.
We do this  expanding  $\Gamma(s)$   as
\beq
\label{a6}
\Gamma (s) = \Gamma (1,s) + \sum_{p=0}^{\infty}
{b_{2p}\over (s+2p)} ,
\eeq
where $\Gamma(z,s)$ is the incomplete gamma function, and
the coefficients of the expansion
are given by $b_{2p} = {(-1)^{2p}/ p!}$.
Using (\ref{a6}) we have:
\beq
\Gamma(s) \mu^{-2s}  = \sum_{p=0}^\infty b_{2p} {\mu^{4p}
\over s+2p} +
\sum_{p=0}^\infty h_p(\mu,s) + \Gamma(1,s) \mu^{-2s} ,
\label{a7}
\eeq
where we have defined
\beq
h_p(\mu,s) = b_{2p} {\mu^{-2s} - \mu^{4p} \over s+2p}.
\eeq
Eq. (\ref{a7}) allows us to write
\beq
\zeta(s) =  {1\over \Gamma(s)} \lim_{\mu \rightarrow 0}
\left\{ \sum_{p=0}^\infty {\tilde{C}_p(\mu)\over s-D_2/2 +p}
+f(\mu,s)\right\},
\label{a8}
\eeq
where
\begin{equation}
    \label{fmu}
    f(\mu,s) = g(\mu,s) -\sum_{p=0}^{\infty} h_p(\mu,s) -
    \Gamma(1,s) \mu^{-2s},
\end{equation}
and the {\it modified} coefficients $\tilde{C}_p(\mu)$ are
then given  by
\beq
\tilde{C}_{D_2/2+2p}(\mu) = C_{D_2/2+2p}(\mu) -
b_{2p}\mu^{4p}.
\eeq
We note that the only coefficients which are modified are
$C_{D_2/2}, C_{D_2/2+1}, C_{D_2/2+2}, \dots$.
Taking the limit $\mu\to0$, we obtain the  main result of
this appendix
\begin{equation}
    \label{mittagleffler}
    \zeta(s) = {1\over \Gamma(s)}
    \left\{ \sum_{p=0}^\infty {\tilde{C}_p\over s-D_2/2 +p}
        +f(s)\right\},
\end{equation}
with
\begin{equation}
    \label{tildecp}
    \tilde{C}_{p}\equiv\lim_{\mu\to0}
    \tilde{C}_p(\mu)=C_p(0)-\delta_{2 p, D_2},
\end{equation}
where $\delta_{p,p'}$ is the Kronecker delta, the limit of
$C_p(\mu)$ can be taken because they are polynomials in
$\mu$, and the function $f(s)= f(0,s) $ is analytic and
finite by construction, although it can be explicitly
checked from (\ref{fmu}).

In conclusion, the result
(\ref{mittagleffler},\ref{tildecp}) implies that for a
laplacian with one zero eigenvalue, there also exists a
Mittag-Leffer-like expansion for the 'primed' zeta function
(\ref{a1},\ref{zeta}),  changing only the Seeley-DeWitt coefficient
$C_{D_2/2}$ by $C_{D_2/2}-1$. That is why we can consider
Eq. (\ref{mtmassive}) valid in general (with either $g_0=1$
or $0$), replacing $C_{D_2/2}$ by $C_{D_2/2}-g_0$.

It is now easy to expand around $s=p$ and a simple
calculation gives
\begin{equation}
    \label{ml}
    \Gamma(s)\zeta(s) \left. \right|_{s=p}= {\tilde
C_{D_2/2-p}\over s-p} +\Omega_p+ {\cal O}((s-p)^2),
\end{equation}
with
\begin{equation}
    \label{omegap}
    \Omega_p\equiv\sum_{p' \neq D_2/2-p} {\tilde C_{p'}\over
      p+p'-D_2/2} + f(p).
\end{equation}

\section{$(D+1)\to 5$ Reduction}
\label{app:5dreduct}

In section \ref{scales}, we have argued that there
exists a range of energies where the theory is
effectively 5 dimensional, as illustrated in
Fig. 2.

In this appendix we show the dimensional reduction
procedure from $(D+1)$ dimensions down to 5 dimensions,
which allows contact with the language of
\cite{pujolas2}.

The reduction from the higher dimensional theory
(\ref{modelaction}) to 5 dimensions is performed by the
compactification on the internal manifold $\Sigma$. This amounts
to keeping only the $\Sigma$-zero modes of the fields defined in
$(D+1)$ dimensions.

In this section we denote
collectively the four dimensional Minkowski coordinates
$x^\mu$ and the orbifold $x^5$
by $x^a$. 

For the sake of simplicity,
we shall consider only the breathing mode of $\Sigma$
in the internal
components of the metric.
As well, we shall freeze
the $\{a,i\}$ components
(the graviphotons) to zero.

Thus, the
ansatz for the metric that we shall adopt depends
on the internal coordinates $X^i$ only through the
background geometry on $\Sigma$, and on $x^a$ through
the five dimensional graviton $g^{(5)}_{ab}$, and a
dilaton $\sigma$,
\begin{equation}
    \label{5dreduct}
    ds^2=g^{(5)}_{ab}(x^c)dx^a dx^b+
    R^2 e^{2\sigma(x^c)} \gamma_{ij}dX^idX^j ~.
\end{equation}

As for the sigma model scalars,
we shall also freeze them to
their value in the background,
$\phi^a=\phi^a(X^i)$.

The action (\ref{modelaction}) corresponding to
this ansatz is
\begin{eqnarray}
   S_5 = - v_{{}_\Sigma} R^{D_2} &\Biggl[&
   \int d^{5}x \sqrt{g_{(5)}}
    e^{D_2 \sigma}
    \left\{ M^{D-1} \left( {\cal R}_{(5)}
        - D_2(D_2-1) (\partial \sigma)_{(5)}^2 \right)
        + \Lambda\right\}\cr
    &+&\int d^{4}x  \sqrt{g_{(5)+}} e^{D_2 \sigma} \; \tau_+
    + \int d^{4}x  \sqrt{g_{(5)-}} e^{D_2 \sigma} \; \tau_-\Biggr],
\end{eqnarray}
where $g^{(5)\pm}_{\mu\nu}$  denote the  metrics on the
branes induced by $g^{(5)}_{ab}$ and
we have performed  the $X$ integration.
We can rewrite this action in the (5 dimensional)
Einstein frame, given by
$g^{{}_E}_{ab}=e^{2D_2\sigma/3} g^{(5)}_{ab}$,
\begin{eqnarray}
\label{5deinst}
S_5=&-& M_{5}^3 \int d^5x \sqrt{g_{{}_E}}\left\{
{\cal R}_{{}_E}+{1\over2}(\partial \phi)_{{}_E}^2
+\Lambda_{5} e^{c\phi} \right\}\\
&-&\int d^{4}x  \sqrt{g_{{}_{E+}}}  \; \tau_{5+} \, e^{c\phi/2}
-\int d^{4}x  \sqrt{g_{{}_{E-}}} \; \tau_{5-}\,e^{c\phi/2}
\end{eqnarray}
where
$$
c^2={2\over3} {D_2\over D_2+3},
$$
the canonical scalar field is
$\phi= -(2D_2/3 c) \sigma $,
$g^{{}_{E\pm}}_{\mu\nu}$  are the  metrics on the
branes induced by $g^{{}_E}_{ab}$, the
5 dimensional Planck mass is given by
$M_{5}^3=v_{{}_\Sigma} R^{D_2} M^{D-1}$, $\Lambda_{5}=M^{1-D}\Lambda$
and  $\tau_{5\pm}=v_{{}_\Sigma} R^{D_2} \tau_\pm$.

The action (\ref{5deinst}) coincides with
the 5 dimensional scalar-tensor model considered
in  \cite{pujolas2}. It was found there that this model
has a solution with a power-law warp factor of the form
\begin{eqnarray}
\label{5dsol}
ds^2_{{}_E}&=&a_{{}_E}^2(z)
\left(dz^2+\eta_{\mu\nu} dx^\mu dx^\nu\right) ,\cr
\phi_0(z)&=&-{\sqrt{6\beta(\beta+1) }}\, \ln (z/z_0)
        \qquad\text{with}\qquad
        a_{{}_E}(z)=\left(z/z_0\right)^{\beta}
\end{eqnarray}
with $\beta=2/(3c^2-2)=-(D_2+3)/3$.  \footnote{In terms of
  the proper coordinate (in the 5 dimensional Einstein
  frame) $y_{{}_E}\propto z^{\beta+1}$ ,
  $a_{{}_E}(y_{{}_E})=(y_{{}_E}/y_{{}_0})^q$ with
  $q=2/3c^2=(D_2+3)/D_2$.}

The brane operators induced by quantum effects on this
background are given by positive powers of
the extrinsic curvature scale (see {\em e.g.} \cite{pujolas2})
${\cal K}_{{}_E\pm}=\beta/z_\pm a_{{}_E\pm}=\beta z_\pm^{-(\beta+1)}$,
\begin{equation}
        \label{5dlocalop}
        \int d^4x \sqrt{g_{{}_E\pm}} {\cal K}_{{}_E\pm}^n
        =\int d^4x \left({z_\pm\over z_0}\right)^{ (4-n) \beta }
                {1\over z_\pm^{n}}
        \propto\int d^4x \; r_\pm^{4+(4 /3) D_2 - (n/3) D_2} ,
\end{equation}
where $n=1,2,3\dots$, we used that the conformal
coordinate $z=e^{k y}$ is the same
in 5 and in $(D+1)$ dimensions, and $a_\pm=1/k z_\pm$.
On the other hand,
we have seen in section \ref{ren}
that the operators generated by the effective
potential due to bulk fields in the model
(\ref{modelaction}) are of the form
\begin{equation}
        \label{D+1localop}
        \int d^4x \sqrt{g_\pm} {\cal R}_{\pm}^N
        \propto \int d^4x  r_\pm^{4+D_2 - 2N}
\end{equation}
where ${\cal R}_{\pm}$ is the intrinsic curvature computed
with the induced metrics on each brane, $g^\pm_{\mu\nu}$.
Here $N=0,1,\dots ,[(D+1)/2]$, and $[~]$ denotes the
integer  part. Now we can identify that
these operators correspond to a number of
powers  of the extrinsic
curvature operator (\ref{5dlocalop}) given by
$$
n={6\over D_2} \; N +1.
$$
We note that all the induced operators can be cast as
powers of the extrinsic curvature for $D_2=1,2,3$ and $6$
only, having in the $D_2=6$ case a one-to-one
correspondence.  For any other value of $D_2$, there exist
higher dimensional local operators that are not simply
powers of ${\cal K}_{{}_E\pm}$, but of some power of $e^\phi$ in
the 5 dimensional effective theory (\ref{5deinst}).

As well, Ref. \cite{pujolas2}
raised the question that the path integral measure of
a bulk scalar field in the effective five dimensional theory (\ref{5deinst})
quantized on the warped vacuum configuration (\ref{5dsol})
is ambiguously defined.
The nontrivial profile of the scalar $\phi$ permits to define
many different conformal frames, all of them equivalent at
the classical level. However, the path integral measure can be defined covariantly
with respect to any of them.
It turns out that the term proportional to
$$
{\ln z_+\over z_+^4} + {\ln z_-\over z_-^4}
$$
in the potential depends on this choice. Several arguments
can be given in favor of possible 'preferred' frames.
For instance, with a measure covariant with respect to the 5 dimensional
Einstein frame metric $g^{{}_E}_{ab}$,
this term is present. But if one chooses covariance
with respect to $g^{(5)}_{ab}$ , there is no such term.
However, in the model presented here, there is no ambiguity in the choice of the measure
since in the $D+1$ dimensional theory there is no scalar with nontrivial profile
along the orbifold.
In the computation presented here, the choice of
the measure shows up (see \cite{pujolas2})
when we subtract the divergences  Eq. (\ref{counterterms}),
covariant precisely with respect to the higher dimensional Einstein frame metric
$g^{(D+1)}_{MN}$.
As a result, when we take into account both the 5 dimensional modes
(the $\Sigma$ KK zero mode)
together with the $D+1$ dimensional ones
(the KK modes excited along the $\Sigma$ as well), we have found that
there is a remaining contribution of this form, see Eqns.
(\ref{finite},\ref{betas},\ref{gammas}).
Anyhow, it should be noted that these
Coleman-Weinberg-like terms do not play a very relevant role in stabilizing of the
moduli.

\vfill \eject

\end{document}